\pdfoutput=1
\documentclass[submission,Phys]{article}

\usepackage{hyperref}
\usepackage{eso-pic} 
 \usepackage{amssymb}
\usepackage{amsfonts}
\usepackage{amsmath}
 \usepackage{xcolor}
\usepackage{csquotes}
 \usepackage{graphicx}
\usepackage{slashed}
 \usepackage{caption}
\usepackage{subcaption}
\usepackage{bm,bbm,bbold}
\usepackage{comment}
\usepackage{cite}

\hypersetup
{
	colorlinks = true,
}

\allowdisplaybreaks

\def\bea{\begin{eqnarray}}
\def\eea{\end{eqnarray}}

\def\ba{\begin{array}}
	\def\ea{\end{array}}

\def\Im{{\rm Im}}

\def\be{\begin{equation}}       \def\ee{\end{equation}}
\def\bea{\begin{eqnarray}}      \def\eea{\end{eqnarray}}
\def\ba{\begin{array}}
	\def\ea{\end{array}}
\def\bnum{\begin{enumerate} }
	\def\enum{\end{enumerate}}

\def\=>{\Rightarrow}
\def\>{\rightarrow}

\def\eye2{Fathbb{I}}

\def\vev#1{\left\langle #1 \right\rangle}

\usepackage{chngcntr}
\counterwithin*{equation}{section}

\def\corr#1{{#1}}

\title{\bf Weak rates in strongly coupled cold quark matter}

\author{Carlos Hoyos${}^1$\footnote{hoyoscarlos@uniovi.es}, Andrea Olzi${}^{2}$\footnote{andrea.olzi@unifi.it},
David Rodriguez-Fernandez${}^3$\footnote{david.rfernandez@upm.es}}

\begin{document}
\maketitle 
\vspace{-22pt}
\begin{center}
\begin{minipage}{0.75\textwidth}
\it{\small 
\makebox[\linewidth][s]{${}^1$Department of Physics and Instituto de Ciencias and Tecnolog\'{\i}as Espaciales de Asturias (ICTEA)}
\makebox[\linewidth][l]{\ Universidad de Oviedo, c/ Leopoldo Calvo Sotelo 18, ES-33007 Oviedo, Spain}
\makebox[\linewidth][s]{${}^2$INFN, Sezione di Firenze and Dipartimento di Fisica e Astronomia, Università di Firenze,}
\makebox[\linewidth][l]{\ Via G. Sansone 1,I-50019 Sesto Fiorentino (Firenze), Italy.}
\makebox[\linewidth][s]{${}^3$Departamento de matemática aplicada a las TIC, Universidad Politécnica de Madrid}
\makebox[\linewidth][l]{\ Nikola Tesla, s/n, ES-28031 Madrid, Spain}
}
\end{minipage}
\end{center}
\vspace{15pt}

\begin{abstract}
The rates of flavor-changing weak processes are crucial in determining the conditions of beta equilibrium in neutron stars and mergers, influencing the damping of oscillations, the stability of rotating pulsars, and the emission of gravitational waves. We derive a formula for these rates at nonzero temperature, to leading order in the Fermi coupling and exact in the QCD coupling. Utilizing a simple phenomenological holographic model dual to QCD, we study massless unpaired quark matter at high densities. We numerically compute the rate for small deviations from beta equilibrium and derive an analytic approximation for small temperatures. Our findings reveal that, compared to the perturbative result, the rate is suppressed by logarithmic factors of the temperature.
\end{abstract}

\newpage

\tableofcontents

\section{Introduction}

Weak interactions play a crucial role in determining the properties of neutron stars and their binary mergers. The equation of state, which dictates the mass and radius of the star, is constrained by the condition of beta equilibrium. This condition establishes relationships among the relative amounts of different quark (or baryon) and lepton species. In mergers, the beta equilibrium condition may be modified due to the temperature dependence of the rates \cite{Alford:2018lhf,Alford:2021ogv,Alford:2023gxq}. When a perturbation in the density drives the matter of the star out of beta equilibrium, weak processes work to restore it and dampen the perturbation. This results in an effective bulk viscosity that is significantly enhanced when the characteristic time of the perturbation resonates with the timescale determined by the rate of weak processes and the density in the star \cite{Sawyer:1989dp,Haensel:1992zz,Haensel:2000vz,Sad:2006egl,Sad:2007afd,Alford:2006gy,Alford:2008pb,Alford:2019qtm,Alford:2019kdw,Alford:2023gxq,Yang:2023ogo} (see \cite{Schmitt:2017efp} for a review of transport in neutron stars).

Oscillations of a rotating neutron star produce gravitational waves that cause the star to lose angular momentum and spin down, a process that is more pronounced for rapidly rotating stars. Despite this, fast-rotating pulsars with periods on the order of milliseconds are observed within a limited range of temperatures. The effective bulk viscosity induced by beta equilibration is likely necessary to explain the existence and location of this stability window \cite{Alford:2010fd,Alford:2013pma,Alford:2015gna}. This viscosity also affects the continuous emission of gravitational waves by a quiescent star \cite{Sieniawska:2019hmd} and could impact the gravitational wave signal produced during the inspiral \cite{Ripley:2023lsq} and post-merger evolution of a neutron star collision \cite{Alford:2017rxf,Radice:2021jtw,Hammond:2022uua,Most:2022yhe,Alford:2022ufz,Chabanov:2023blf,Chabanov:2023abq}.

Although most matter in a neutron star and the remnant of a merger is expected to be primarily composed of hadrons, there are indirect hints of quark matter in the core of heavy neutron stars \cite{Annala:2019puf,Annala:2021gom,Annala:2023cwx}. Simulations of binary mergers suggest that pockets of quark matter may be produced just after the collision \cite{Prakash:2021wpz,Tootle:2022pvd}. Therefore, it is interesting to evaluate the bulk viscosity of quark matter induced by weak processes. At the densities involved, quantum chromodynamics (QCD) cannot be treated perturbatively, and lattice QCD suffers from the sign problem, necessitating other non-perturbative approaches. Holographic models can circumvent these issues and have been applied to the study of dense matter in neutron stars (see \cite{Hoyos:2021uff,Jarvinen:2021jbd} for comprehensive reviews). Specifically, the bulk viscosity induced by weak processes was computed using holographic models in \cite{CruzRojas:2024etx}\footnote{The holographic estimate of the QCD contribution to the bulk viscosity was computed previously and found to be much smaller than the expected weak contribution \cite{Czajka:2018bod,Hoyos:2020hmq,Hoyos:2021njg}.}. However, the bulk viscosity formula involves weak reaction rates that have been computed only perturbatively, so it is not a completely self-contained approach. Furthermore, the existing formulas for weak rates are obtained through tree-level scattering matrix elements of quarks and leptons \cite{Heiselberg:1986pg,Heiselberg:1992bd,Madsen:1993xx}. Since the observables obtained from gauge/gravity duality are correlators of gauge-invariant operators, there is no straightforward way to obtain the rates by extrapolating the tree-level formulas. Unfortunately, a main feature of the bulk viscosity, namely its temperature dependence, is highly sensitive to the behavior of the rate. Thus, the present estimation of the bulk viscosity is hindered by significant uncertainty in this respect.

As a step towards improving the bulk viscosity estimates, we will provide a non-perturbative formula for the rates based on correlators of gauge-invariant operators. We will then use it to obtain the temperature dependence of the rates in a simple holographic model of unpaired quark matter. The model, a phenomenological holographic dual description of QCD \cite{Erlich:2005qh}, has been used extensively, with recent related works including the description of nuclear matter in neutron stars and mergers \cite{Bartolini:2022rkl} and the calculation of neutrino emissivities from flavor current correlators \cite{Jarvinen:2023xrx}. Our goal is not to obtain completely realistic values for the rates, for which more refined models would be needed, but to ascertain possible qualitative changes in the temperature dependence at strong coupling.

Holographic models, despite their utility, come with inherent limitations. They are typically constructed as approximations that capture certain features of QCD, but they do not encompass all the complexities of the theory. One significant limitation is the lack of a first-principles derivation from QCD, leading to uncertainties in their applicability to specific phenomena. Furthermore, these models often rely on large-$N_c$ (number of colors) approximations, which may not accurately reflect the behavior of QCD at the finite $N_c$ relevant to real-world scenarios. The models also face challenges in accurately representing the dynamics at low temperatures and densities where confinement and chiral symmetry breaking are essential. Therefore, while holographic models provide valuable insights and a non-perturbative framework, their predictions should be interpreted with caution and validated against other approaches and experimental data.

Keeping the limitations above in mind, for direct application to the description of neutron stars and their binary mergers, the rate should be computed in the model used to describe matter in the star. The most employed holographic models so far have been those inspired by top-down brane intersections, namely the D3-D7 \cite{Karch:2002sh} and Witten-Sakai-Sugimoto models \cite{Sakai:2004cn,Sakai:2005yt}, and a bottom-up model designed to fit lattice QCD data and other known observables, the V-QCD model \cite{Jarvinen:2011qe}. The D3-D7 model is the simplest but is limited to describing quark matter in the star \cite{Aleixo:2023lue,Hoyos:2016zke,Annala:2017tqz,BitaghsirFadafan:2019ofb,BitaghsirFadafan:2020otb} and must be combined with other nuclear matter models to construct hybrid equations of state for the matter inside the neutron star. The V-QCD model can describe confined matter but is used primarily at large densities, requiring a hybrid approach to describe matter at lower densities outside the core of the star \cite{Jokela:2018ers,Ecker:2019xrw,Jokela:2020piw,Demircik:2021zll,Jokela:2021vwy}. Hybrid equations of state have also been constructed with the Sakai-Sugimoto model \cite{Pinkanjanarod:2020mgi,Burikham:2021xpn,Bartolini:2022gdf,Bartolini:2023wis}, but there are proposals to describe nuclear matter in neutron stars at all densities purely holographically \cite{Zhang:2019tqd,Kovensky:2021ddl,Kovensky:2021kzl}. The Sakai-Sugimoto model has also been used to describe more exotic quark stars \cite{Burikham:2010sw}. Other proposals to describe dense matter in neutron stars include Einstein-Maxwell-Dilaton theories and other holographic phenomenological models \cite{Ghoroku:2021fos,Mamani:2020pks,Zhang:2022uin}. Computing the weak rate and bulk viscosity to estimate the location of the stability window of rotating pulsars would serve as an additional constraint on these models, as well as on any other phenomenological model for which such a calculation is possible.

The outline of the paper is as follows. In Section~\ref{sec:qft}, we introduce the weak reaction rates for beta equilibrium of quark matter and derive a non-perturbative formula in terms of correlators of flavor currents. Next, in Section~\ref{sec:holo}, we use a simple holographic model to compute flavor current correlators in a high-density state and study their dependence on frequency and spatial momentum. Finally, in Section~\ref{sec:rate}, we combine the results of the previous sections to produce analytical and numerical estimates of the weak rates at low temperatures. We compare these to the perturbative results and discuss potential improvements in Section~\ref{sec:conc}. Technical details regarding the calculation of correlators in the holographic model are provided in the appendices.

\section{Weak reaction rates}\label{sec:qft}

In this section, we derive non-perturbative formulas in the coupling of the strong interactions for the weak reaction rates of quark matter. The formulas correspond to the leading contribution of the electroweak interactions, whose corrections could in principle be computed perturbatively. We also assume that the times and length scales involved are much larger than the scale set by the inverse of the $W$ boson mass.

For the densities expected in the interior of neutron stars and mergers, one should be able to restrict to three quark flavors ($q=u,d,s$) and two lepton flavors ($\ell=e,\mu$). The relevant weak processes, or reactions, are produced by a $W$-boson exchange or emission:
\begin{equation}\label{eq:weakreac}
u+d\rightleftharpoons u+s,\quad u+\ell^- \to  d,s+\nu_{\ell}, \quad d,s \to u+\ell^- +\bar{\nu}_{\ell},\quad \ell_1^- \to  \ell_2^-+\nu_{\ell_1}+\bar{\nu}_{\ell_2}\;.
\end{equation}
Assuming neutrinos are not trapped, the beta-equilibrium conditions in quark matter implied by these reactions are
\begin{equation}\label{eq:betaeq}
\mu_u+\mu_\ell=\mu_d=\mu_s,\qquad \mu_\ell=\mu_e=\mu_\mu\,.
\end{equation}
\corr{These relations assume that the different fermion species are at thermal equilibrium. However, that is not the case for neutrinos and for large enough temperatures, the relations among different chemical potentials might be shifted. This was analyzed for baryonic matter in \cite{Alford:2018lhf,Alford:2021ogv,Alford:2023gxq}.}

Even if quarks are confined to hadrons the same reactions will change their type among protons, neutrons, and hyperons. In this situation, the beta equilibrium conditions should be given in terms of baryon chemical potentials. Note that in order for any of the species of fermions to be present it is necessary that the chemical potential is larger than the mass of the corresponding particle.

If a perturbation takes the chemical potentials out of their equilibrium values, the weak reactions will not be in balance, and the densities will change according to the weak rates $\Gamma$:\footnote{We have omitted the label on the neutrinos when there is just one lepton involved and put the same numerical sub-index as the corresponding lepton when there are two.}  
\begin{subequations}\label{eq:ratedef}
\begin{eqnarray}
\frac{d n_u}{dt} &=&\sum_{\ell=e,\mu}\left( \Gamma_{d\to u\ell\bar{\nu}}-\Gamma_{u\ell\to d\nu}+ \Gamma_{s\to u \ell\bar{\nu}}-\Gamma_{u\ell\to s\nu}\right)\;,\\
\frac{d n_d}{dt} &=&\Gamma_{su\to ud}-\Gamma_{ud\to su}+\sum_{\ell=e,\mu}\left( \Gamma_{u\ell\to d\nu}-\Gamma_{d\to u\ell\bar{\nu}}\right)\;,\\
\frac{d n_s}{dt} &=&\Gamma_{ud\to su}-\Gamma_{su\to ud}+ \sum_{\ell=e,\mu}\left(\Gamma_{u\ell\to s\nu}-\Gamma_{s\to u \ell\bar{\nu}}\right)\;,\\
\frac{d n_{\ell_1}}{dt} &=&\Gamma_{d\to u\ell_1\bar{\nu}}-\Gamma_{u\ell_1\to d\nu}+\Gamma_{s\to u \ell_1\bar{\nu}}-\Gamma_{u\ell_1\to s\nu}+\Gamma_{\ell_2\to\ell_1\nu_2\bar{\nu}_1}-\Gamma_{\ell_1\to \ell_2\nu_1\bar{\nu}_2}\;.
\end{eqnarray}
\end{subequations}
It is easy to confirm that the total baryon number and electric charge are conserved by these processes
\begin{equation}
n_B=\frac{1}{3}(n_u+n_d+n_s),\quad n_Q=\frac{2}{3}n_u-\frac{1}{3}n_d-\frac{1}{3}n_s-n_e-n_\mu,\qquad \frac{dn_B}{dt}=\frac{dn_Q}{dt}=0\,.
\end{equation}

Assuming the deviation of the chemical potentials from their beta-equilibrium values $\delta \mu_{f^a}$ is small compared to the baryon chemical potential $\mu_B=\mu_u+\mu_d+\mu_s$, we can expand the rates to linear order in the differences
\begin{subequations}\label{eq:ratecoefdef}
\begin{eqnarray}
\Gamma_{ud\to su}-\Gamma_{su\to ud} &\approx & \lambda_{ds} (\delta \mu_s-\delta \mu_d),\\
\Gamma_{u\ell\to d\nu}-\Gamma_{d\to u\ell\bar{\nu}} &\approx & \lambda_{ud}^\ell (\delta\mu_d-\delta \mu_u-\delta \mu_{\ell}),\\
\Gamma_{u\ell\to s\nu}-\Gamma_{s\to u\ell\bar{\nu}} &\approx & \lambda_{us}^\ell (\delta \mu_s-\delta \mu_u-\delta \mu_{\ell}),\\
\Gamma_{e\to\mu\nu_2\bar{\nu}_1}-\Gamma_{\mu\to e\nu_1\bar{\nu}_2} &\approx & \lambda_{e\mu} (\delta \mu_{\mu}-\delta \mu_{e}).
\end{eqnarray}
\end{subequations}

In some cases, for instance when quark matter is close to being symmetric, the lepton chemical potential could be below the muon mass $\mu_\ell<m_\mu$. When this happens, muons quickly decay and the only leptons present are electrons. The equilibration equations simplify somewhat
\begin{subequations}\label{eq:ratedefonlye}
\begin{eqnarray}
\frac{d n_u}{dt} &=& \Gamma_{d\to u e\bar{\nu}}-\Gamma_{ue\to d\nu}+ \Gamma_{s\to u e\bar{\nu}}-\Gamma_{u e\to s\nu}\;,\\
\frac{d n_d}{dt} &=&\Gamma_{su\to ud}-\Gamma_{ud\to su}+\Gamma_{u e\to d\nu}-\Gamma_{d\to u e\bar{\nu}}\;,\\
\frac{d n_s}{dt} &=&\Gamma_{ud\to su}-\Gamma_{su\to ud}+ \Gamma_{u e\to s\nu}-\Gamma_{s\to u e\bar{\nu}}\;,\\
\frac{d n_e}{dt} &=&\Gamma_{d\to ue\bar{\nu}}-\Gamma_{ue\to d\nu}+\Gamma_{s\to u e\bar{\nu}}-\Gamma_{ue\to s\nu}\;.
\end{eqnarray}
\end{subequations}
For small deviations from beta-equilibrium, the rates are expanded as \eqref{eq:ratecoefdef}, but taking into account only the electron contribution. Moreover, the coefficient $\lambda_{e\mu}$ does not involve strong coupling physics, so we will ignore it in the following, as it can be computed using the standard methods. Our strategy to derive the other coefficients is to arrive at an expression like \eqref{eq:ratedef} from Ward identities capturing the non-conservation of flavor currents due to the coupling to the weak sector. 

\subsection{Flavor Ward identities}

The theory of strong interactions, QCD, respects flavor symmetries so that there are no flavor-changing processes induced by them. Weak interactions on the other hand are not flavor symmetric, and processes such as those in \eqref{eq:weakreac} are possible. At energy scales much below the $W$ mass, the flavor-changing processes are captured by adding Fermi's interaction term to the effective action
\begin{equation}\label{eq:FermiAction}
{\cal L}_{\rm Fermi}=-2\sqrt{2} G_F (J_{ch}^\mu)^\dagger J_{ch\,\mu},
\end{equation}
where the $J_{ch}^\mu$ is the left-charged current
\begin{equation}
J_{ch}^\mu=\overline{\nu}_{e\,L} \gamma^\mu e_L+\overline{\nu}_{\mu\,L} \gamma^\mu \mu_L+\cos \theta_C\, \overline{u}_L\gamma^\mu d_L+\sin \theta_C\, \overline{u}_L\gamma^\mu s_L,\ \
\end{equation}
with $G_F$ Fermi's constant and $\theta_C$ the Cabibbo angle giving the change of basis between quark mass and flavor eigenstates. In addition to Fermi's term, flavor symmetry can also be broken by flavor-preserving mass terms and electromagnetic and weak neutral interactions.

In the absence of any flavor-breaking terms, there would be a $SU(3)_L\times SU(3)_R$ global symmetry for flavor rotations among quarks and a $SU(4)_L\times SU(4)_R$ flavor global symmetry for leptons. To each of these symmetries, we can associate a conserved current $J^\mu_{f\, \chi}$, where $f=l,q$ for leptons or quarks and $\chi=L,R$ for left or right chirality. Since the symmetry is non-Abelian the current components transform under flavor rotations. Denoting by $T^A_{f\,\chi}$ the generators of the flavor symmetry, the transformation of the current is
\begin{equation}
\delta_{\theta_{f\, \chi}} J^\mu_{f\, \chi}=i\theta_{f\,\chi}^A[J^\mu_{f\, \chi},T^A_{f\,\chi}]\;.
\end{equation}
We will take the usual normalization for the generators ${\rm tr}(T_{f\,\chi}^A T_{f\,\chi}^B)=\frac{1}{2}\delta^{AB}$. 

A vector flavor transformation corresponds to a transformation with equal angles for both chiralities $\theta_{f\,L}^A=\theta_{f\,R}^A=\theta_f^A$. In the following vector currents and generators will be denoted as $J^\mu_f$ and $T^A_f$. If there are flavor-symmetry breaking terms in the action ${\cal L}_{\rm br}$, the Ward identity associated with a vector transformation becomes
\begin{equation}
2\theta_f^A{\rm tr}\left(\partial_\mu J^\mu_f\,T^A_f\right)=-\delta_{\theta^A_{f\,L}}{\cal L}_{\rm br}-\delta_{\theta^A_{f\,R}}{\cal L}_{\rm br}\Big|_{\theta_{f\,L}^A=\theta_{f\,R}^A=\theta_f^A}.
\end{equation}
In the following, it will be convenient to introduce the notation for the currents
\begin{equation}
(J_{f\,\chi}^\mu)^a_{\ b}=\bar{f}_{\chi\,b} \gamma^\mu f_\chi^a\,,\qquad f^a=q^a,l^a;\quad q^a=u,d,s;\; l^a=\nu_e,e,\nu_\mu,\mu.
\end{equation}

We are interested in the flavor-changing contributions introduced by Fermi's interaction. We can simplify the analysis by noticing that since Fermi's interaction is symmetric under a rotation between the two lepton generations $(\nu_e,e)\leftrightarrow (\nu_\mu,\mu)$, we can actually restrict lepton transformations to $SU(2)\times SU(2)\subset SU(4)$, with each $SU(2)$ rotating a charged lepton and its corresponding neutrino within each generation. We leave details of the calculation, which is straightforward but tedious, to Appendix~\ref{app:flavorbreak}. The Ward identities for the diagonal components of the vector flavor currents are
\begin{subequations}\label{eq:wardident}
\begin{eqnarray}
\notag \partial_\mu \vev{(J_q^\mu)^u_{\ u}} &= & i\sqrt{2}G_F \eta_{\mu\nu}\sum_{\ell=e,\mu}\left[\cos\theta_C\vev{(J_{q\,L}^\mu)^u_{\ d}(J_{l\,L}^\nu)^d_{\ u}-(J_{q\,L}^\mu)^d_{\ u}(J_{l\,L}^\nu)^{\nu_\ell}_{\ \ell}}\right.\\ 
&\quad & \left. +\sin\theta_C\vev{(J_{q\,L}^\mu)^u_{\ s}(J_{l\,L}^\nu)^\ell_{\ \nu_\ell}-(J_{q\,L}^\mu)^s_{\ u}(J_{l\,L}^\nu)^{\nu_\ell}_{\ \ell}} \right],\\
\notag \partial_\mu \vev{(J_q^\mu)^d_{\ d}} & =& i\sqrt{2} G_F\eta_{\mu\nu}\left[\cos\theta_C \sin\theta_C\eta_{\mu\nu}\vev{(J_{q\, L}^\mu)^u_{\ s} (J_{q\,L}^\nu)^d_{\ u}-(J_{q\,L}^\mu)^s_{\ u}(J_{q\,L}^\nu)^u_{\ d} }\right.\\
&\quad & \left.-\cos\theta_C\sum_{\ell=e,\mu}\vev{(J_{q\,L}^\mu)^u_{\ d}(J_{l\,L}^\nu)^\ell_{\ \nu_\ell}-(J_{q\,L}^\mu)^d_{\ u}(J_{l\,L}^\nu)^{\nu_\ell}_{\ \ell}}\right],\\
\notag \partial_\mu \vev{(J_q^\mu)^s_{\ s}} &=& i\sqrt{2} G_F\eta_{\mu\nu}\left[-\cos\theta_C \sin\theta_C\eta_{\mu\nu}\vev{(J_{q\, L}^\mu)^u_{\ s} (J_{q\,L}^\nu)^d_{\ u}-(J_{q\,L}^\mu)^s_{\ u}(J_{q\,L}^\nu)^u_{\ d} }\right.\\
&\quad & \left.-\sin\theta_C\sum_{\ell=e,\mu}\vev{(J_{q\,L}^\mu)^u_{\ s}(J_{l\,L}^\nu)^\ell_{\ \nu_\ell}-(J_{q\,L}^\mu)^s_{\ u}(J_{l\,L}^\nu)^{\nu_\ell}_{\ \ell}}\right],\\
\notag \partial_\mu \vev{(J_l^\mu)^{\nu_\ell}_{\ \nu_\ell}} &=& i\sqrt{2}G_F \eta_{\mu\nu}\left[-\cos\theta_C\vev{(J_{q\,L}^\mu)^u_{\ d}(J_{l\,L}^\nu)^\ell_{\ \nu_\ell}-(J_{q\,L}^\mu)^d_{\ u}(J_{l\,L}^\nu)^\ell_{\ \nu_\ell}}\right.\\
\notag &\quad & \left. -\sin\theta_C\vev{(J_{q\,L}^\mu)^u_{\ s}(J_{l\,L}^\nu)^\ell_{\ \nu_\ell}-(J_{q\,L}^\mu)^s_{\ u}(J_{l\,L}^\nu)^{\nu_\ell}_{\ \ell}}\right.\\
&\quad & \left.-\vev{(J_{l\,L}^\mu)^{\nu_{\ell'}}_{\ \ell'}(J_{l\,L}^\nu)^\ell_{\ \nu_\ell}-(J_{l\,L}^\mu)^{\ell'}_{\ \nu_{\ell'}}(J_{l\,L}^\nu)^\ell_{\ \nu_\ell}} \right],\\
\notag \partial_\mu \vev{(J_l^\mu)^\ell_{\ \ell}} &=& i\sqrt{2}G_F \eta_{\mu\nu}\left[\cos\theta_C\vev{(J_{q\,L}^\mu)^u_{\ d}(J_{l\,L}^\nu)^\ell_{\ \nu_\ell}-(J_{q\,L}^\mu)^d_{\ u}(J_{l\,L}^\nu)^{\nu_\ell}_{\ \ell}}\right.\\
\notag &\quad &\left. +\sin\theta_C\vev{(J_{q\,L}^\mu)^u_{\ s}(J_{l\,L}^\nu)^\ell_{\ \nu_\ell}-(J_{q\,L}^\mu)^s_{\ u}(J_{l\,L}^\nu)^{\nu_\ell}_{\ \ell}}\right.\\
&\quad & \left.+\vev{(J_{l\,L}^\mu)^{\nu_{\ell'}}_{\ \ell'}(J_{l\,L}^\nu)^\ell_{\ \nu_\ell}-(J_{l\,L}^\mu)^{\ell'}_{\ \nu_{\ell'}}(J_{l\,L}^\nu)^{\nu_\ell}_{\ \ell}} \right].
\end{eqnarray}
\end{subequations}
Where $(\ell,\ell')=(e,\mu)$ or $(\mu,e)$. Note that other terms that break the non-Abelian flavor symmetry, but are diagonal in flavor, do not contribute to the Ward identities of the diagonal components. One can easily confirm that the expressions above are consistent with the conservation of baryon and lepton numbers as well as the electric charge:
\begin{equation}
\begin{split}
   & \partial_\mu\vev{(J_q^\mu)^u_{\ u}+(J_q^\mu)^d_{\ d}+(J_q^\mu)^s_{\ s}}=0,\quad \partial_\mu\vev{(J_l^\mu)^{\nu_\ell}_{\ \nu_\ell}+(J_l^\mu)^\ell_{\ \ell}}=0, \\
    &  \partial_\mu\vev{\frac{2}{3}(J_q^\mu)^u_{\ u}-\frac{1}{3}(J_q^\mu)^d_{\ d}-\frac{1}{3}(J_q^\mu)^s_{\ s}-\sum_{\ell=e,\mu} (J_l^\mu)^\ell_{\ \ell}}=0.
    \end{split}
\end{equation}
The time components of the diagonal components of the flavor vector currents equal the corresponding quark or lepton densities
\begin{equation}
n_u=(J_q^0)^u_{\ u},\ n_d=(J_q^0)^d_{\ d},\ n_s=(J_q^0)^s_{\ s},\ n_\ell=(J_l^0)^\ell_{\ \ell}.
\end{equation}
Compared with \eqref{eq:ratedef} for homogeneous configurations, the rates are determined by the expectation value of a non-diagonal operator quadratic in the flavor currents appearing on the right-hand side of the Ward identities. We can think of such a quadratic operator as obtained from a two-point correlator in the (properly regularized) limit of coincident points. In the absence of flavor-breaking terms, the most general form of the flavor current correlators would be
\begin{equation}\label{eq:invcorr}
\vev{(J_{f\, L}^\mu)^a_{\ b} (J_{f\,L}^\nu)^c_{\ d}}= G_1 \delta^a_b\delta^c_d+G_2 \delta^a_d \delta^c_b\,.
\end{equation}
If there are mass terms in the action of the form ${\cal L}_m=\bar{f}_a M^a_{f\ b} f^b$, or chemical potentials introduced through a background flavor gauge field $A^a_ {f\ b}$, there could also be terms where we replace some or all of the deltas in \eqref{eq:invcorr} by $M_f$ or $A_f$, or contractions like $M^a_{f\ c} M^c_{f\ b}$,  $M^a_{f\ c} A^c_{f\ b}$, etc. This potentially can produce many new terms, however, for diagonal mass matrices and chemical potentials, the non-diagonal current correlators appearing on the right-hand side of the Ward identities would still vanish, and similarly for contributions originating from interactions that are diagonal in flavor. Therefore, the non-diagonal current correlators contributing to the Ward identities must be proportional to the flavor-changing coupling $G_F$. We will derive in the following a closed expression for the $O(G_F^2)$ contribution to the rates in thermal field theory.

\subsection{Leading contribution in the Fermi coupling}

We will assume that the theory is at finite temperature \corr{$T=1/\beta$} and use the real-time formalism to compute the current correlators. In general, thermal correlation functions are computed by insertions in a path integral defined along a contour $C$ in complex time. The \corr{path-ordered} two-point function of two operators ${\cal O}_1$ and ${\cal O}_2$ is
\begin{equation}
\vev{T_C\left[{\cal O}_1(x_1){\cal O}_2(x_2)\right]}.
\end{equation}
The complex time path, or Schwinger-Keldysh path, extends along \corr{$C=(t_i,t_f)\cup (t_f,t_f-i\sigma)\cup (t_f-i\sigma,t_i -i\sigma)\cup (t_i-i\sigma,t_i-i\beta)$}, with $\sigma>0$ and  $t_i\to-\infty$, $t_f\to+\infty$. We take the point $x_1$ to be at the first leg of the complex time path on the real axis. Then we have two possible contributions, \corr{depending on where the point $x_2$ is located}. If $x_2$ is on the same leg \corr{as $x_1$}, then the thermal correlator coincides with the usual time-ordered Feynman correlator
\begin{equation}
\vev{T_C\left[{\cal O}_1(x_1){\cal O}_2(x_2)\right]}=\vev{T\left[{\cal O}_1(x_1){\cal O}_2(x_2)\right]}=G^F_{{\cal O}_1{\cal O}_2 }(t_1-t_2,\bm{x_1}-\bm{x_2}).
\end{equation}
If, on the other hand, $x_2$ is on the third leg \corr{of the contour $C$} ($\Im\, t'=-\sigma$), the correlator is the Wightman correlator
\begin{equation}
\vev{T_C\left[{\cal O}_1(x_1){\cal O}_2(x_2)\right]}=\vev{{\cal O}_2(x_2){\cal O}_1(x_1)}=G^<_{{\cal O}_1{\cal O}_2 }(t_1-t_2+i\sigma,\bm{x_1}-\bm{x_2}).
\end{equation}
With this in mind, let us find the effect of Fermi's interaction \eqref{eq:FermiAction} when we treat it perturbatively. The thermal expectation value of an operator ${\cal O}$ will be, to leading order in the Fermi coupling,
\begin{equation}
\vev{{\cal O}(x)}_{G_F}=\vev{T_C\left[{\cal O}(x) e^{i\int_C  {\cal L}_{\rm Fermi}}\right]}_0\approx \vev{{\cal O}(x)}_{0}+i\int_C d^4 x'\vev{T_C\left[{\cal O}(x)  {\cal L}_{\rm Fermi}(x')\right]}_0.
\end{equation}
In the expression above $\vev{X}_0$ indicates that the expectation value of $X$ is computed in the absence of Fermi's interaction. Since $x$ is in the first leg of the complex time contour and  $x'$ runs over its full extent, we have contributions from both the first and third leg \corr{of $C$, that translate into Feynman and Wightman correlators of the operator with the Fermi interaction respectively}
\begin{equation}
\int_C d^4 x'\vev{T_C\left[{\cal O}(x)  {\cal L}_{\rm Fermi}(x')\right]}_0=\int d^4x' \left[G_{{\cal O}{\cal L}_{\rm Fermi}}^F(t-t',\bm{x}-\bm{x'})+G_{{\cal O}{\cal L}_{\rm Fermi}}^<(t-t'+i\sigma,\bm{x}-\bm{x'})\right].
\end{equation}
In the evaluation of the right-hand side of the Ward identities \eqref{eq:wardident}, ${\cal O}$ is quadratic in the flavor currents, as well as ${\cal L}_{\rm Fermi}$. Therefore, the leading $O(G_F^2)$ contribution is proportional to correlators of the form
\begin{equation}
\begin{split}
&\vev{\left[(J_f^\mu)^a_{\ b}(x) (J_{f}^\nu)^c_{\ d}(x)\right]\left[(J_{f'}^\alpha)^{a'}_{\ b'}(x') (J_{f'}^\beta)^{c'}_{\ d'}(x')\right]}_0\approx \\
&\delta_{ff'}\left[\vev{(J_f^\mu)^a_{\ b}(x)(J_f^\alpha)^{a'}_{\ b'}(x')}_0\vev{(J_{f}^\nu)^c_{\ d}(x) (J_f^\beta)^{c'}_{\ d'}(x')}_0+\vev{(J_f^\mu)^a_{\ b}(x) (J_f^\beta)^{c'}_{\ d'}(x')}_0\vev{(J_{f}^\nu)^c_{\ d}(x)(J_f^\alpha)^{a'}_{\ b'}(x')}_0\right].
\end{split}
\end{equation}
The factorization above is an approximation. For holographic models dual to a large-$N_c$ theory it will happen naturally as part of the large-$N_c$ expansion. More generally, the non-factorizable contributions would originate from effective eight-fermion interactions, so it is perhaps not unreasonable to assume that they would introduce relatively small corrections even in the finite-$N_c$ strongly coupled theory. 

As we discussed before, in the absence of the Fermi interaction, non-vanishing current correlators should be diagonal in flavor, thus
\begin{equation}
G^{\mu\alpha}_{f;ab,a'b'}(x-x')\equiv \vev{(J_f^\mu)^a_{\ b}(x)(J_f^\alpha)^{a'}_{\ b'}(x')}_0\neq 0,\quad  \Rightarrow\ \ (a=b\; \land\; a'=b') \ \lor\ (a=b'\; \land\; a'=b).
\end{equation}
This limits the number of non-zero contributions to the four-current thermal correlators. We will now write the leading contributions in the Fermi coupling to the Ward identities \eqref{eq:wardident} in terms of the Fourier transforms of the Feynman and Wightman correlators of the flavor currents
\begin{equation}
  G^{K\,\mu\alpha}_{f;ab,a'b'}(x-x')=\int \frac{d^4 k}{(2\pi)^4}e^{ik\cdot (x-x')} G^{K\,\mu\alpha}_{f;ab,a'b'}(k),\qquad K=F,<\, .  
\end{equation}
It will be convenient to define
\begin{equation}\label{eq:gammadef}
    \gamma_{_{f_1^a f_2^b\to f_1^cf_2^d}}=\gamma^F_{_{f_1^af_2^b\to f_1^c f_2^d}}+\gamma^<_{_{f_1^af_2^b\to f_1^c f_2^ d}},
\end{equation}
with ($K=F,<$)
\begin{equation}\label{eq:gammacorr}
   \gamma_{f_1^a f_2^b\to f_1^c f_2^d}^{K}=\eta_{\mu\nu}\eta_{\alpha\beta}\int \frac{d^4 k}{(2\pi)^4}
  \left[G^{K\,\mu\alpha}_{f_1;ac,ca}(k)G^{K\,\nu\beta}_{f_2;bd,db}(-k)-G^{K\,\mu\alpha}_{f_1;ca,ac}(k)G^{K\,\nu\beta}_{f_2;db,bd}(-k)\right]\,.
\end{equation}
In terms of these quantities, the leading order contributions in the Fermi coupling to the flavor Ward identities are
\begin{subequations}\label{eq:Wardlead}
\begin{eqnarray}
    \partial_\mu \vev{(J_q^\mu)^u_{\ u}} &\approx & -4 G_F^2\sum_{\ell=e,\mu} \left(\cos^2\theta_C \gamma_{u\ell\to d\nu}+\sin^2\theta_C \gamma_{u\ell\to s\nu}\right),\\
    \partial_\mu \vev{(J_q^\mu)^d_{\ d}} &\approx & 4 G_F^2 \left(-\sin\theta_C^2\cos^2\theta_C \gamma_{ud\to su}+\cos^2\theta_C \sum_{\ell=e,\mu}\gamma_{u\ell\to d\nu}\right),\\
     \partial_\mu \vev{(J_q^\mu)^s_{\ s}} &\approx & 4 G_F^2 \left(\sin\theta_C^2\cos^2\theta_C \gamma_{ud\to su}+\sin^2\theta_C \sum_{\ell=e,\mu}\gamma_{u\ell\to s\nu}\right),\\
     \partial_\mu \vev{(J_l^\mu)^{\nu_\ell}_{\ \nu_\ell}} &\approx & 4 G_F^2 \left(\cos^2\theta_C \gamma_{u\ell\to d\nu}+\sin^2\theta_C \gamma_{u\ell\to s\nu}+\gamma_{\ell\nu'\to \nu\ell'}\right),\\
    \partial_\mu \vev{(J_l^\mu)^\ell_{\ \ell}} &\approx & -4 G_F^2 \left(\cos^2\theta_C \gamma_{u\ell\to d\nu}+\sin^2\theta_C \gamma_{u\ell\to s\nu}+\gamma_{\ell\nu'\to \nu\ell'}\right).
\end{eqnarray}
\end{subequations}
Where in the last two rows $(\ell,\ell')=(e,\mu)$ or $(\mu,\ell)$ and $\nu,\nu'$ are the neutrinos corresponding to $\ell,\ell'$ respectively.

\subsection{Formulas for the rates}

A direct comparison of the equilibration equations  \eqref{eq:ratecoefdef} and the Ward identities \eqref{eq:Wardlead} results in the identification of the leading order contribution in the Fermi coupling to the rates as
\begin{subequations}\label{eq:formularates}
\begin{eqnarray}
\Gamma_{u\ell\to d\nu}-\Gamma_{d\to u \ell\bar{\nu}} &\approx & 4 G_F^2\cos^2\theta_C \gamma_{u\ell\to d\nu}\; ,\\
\Gamma_{u \ell\to s\nu}-\Gamma_{s\to u \ell\bar{\nu}} &\approx &  4 G_F^2\sin^2\theta_C \gamma_{u\ell\to s\nu}\; ,\\
\Gamma_{ud \to su}-\Gamma_{su \to ud} &\approx &  4 G_F^2\sin\theta_C^2\cos^2\theta_C \gamma_{ud\to su}\; ,\\
\Gamma_{\ell_1\to \ell_2\nu_1\bar{\nu}_2}-\Gamma_{\ell_2\to \ell_1\nu_2\bar{\nu}_1} &\approx & 4 G_F^2\gamma_{\ell_1\nu_2\to \nu_1\ell_2}\;.
\end{eqnarray}
\end{subequations}
With the factors $\gamma$ defined in \eqref{eq:gammadef} and \eqref{eq:gammacorr} in terms of thermal correlation functions. We can further simplify the expressions by first noting that, if the system has parity and time reversal invariance, the Feynman correlator satisfies
\begin{equation}
      G_{f;ab,ba}^{F\,\mu\nu}(-k_0,-\bm{k})=G_{f;ba,ab}^{F\,\nu\mu}(k_0,\bm{k}).
\end{equation}
A straightforward calculation then shows that $\gamma^F_{_{f_1^af_2^b\to f_1^c f_2^d}}=0$ in this case. The Wightman correlator on the other hand is proportional to the spectral function, defined as the imaginary part of the retarded correlator:
\begin{equation}\label{eq:thermalcorr}
G^{<\,\mu\nu}_{f;ab,ba}(k_0,\bm{k})=n_{f;ab}(k_0)\rho_{f;ab}^{\mu\nu}(k_0,\bm{k}),\quad \rho_{f;ab}^{\mu\nu}=-2\,\Im \left(G^{R\,\mu\nu}_{f;ab,ba}\right),
\end{equation} 
where the factor multiplying the spectral function is the thermal distribution
\begin{equation}
n_{f;ab}(k_0)=\frac{1}{e^{\beta(k_0+\mu_{f^a}-\mu_{f^b})}-1}\,,
\end{equation}
with $\mu_{f^a}$ the chemical potential for the fermion species $f^a$. We now introduce \eqref{eq:thermalcorr} in \eqref{eq:gammacorr}, and use that
\begin{equation}
n_{f;ab}(-k_0)=-(1+n_{f;ba}(k_0)), \qquad \rho_{f;ab}^{\mu\alpha}(-k_0,-\bm{k})=-\rho_{f;ba}^{\alpha\mu}(k_0,\bm{k})\;,
\end{equation}
where the last identity follows from time-reversal invariance and parity. After this, one arrives at
\begin{equation}\label{eq:gamma}
\gamma_{_{f_1^a f_2^b\to f_1^cf_2^d}}=\gamma^<_{_{f_1^af_2^b\to f_1^c f_2^ d}}=\eta_{\mu\nu}\eta_{\alpha\beta}
  \int \frac{d^4 k}{(2\pi)^4}\left[(n_{f_1;ac}(k_0)-n_{f_2;db}(k_0))\rho^{\mu\alpha}_{f_1;ac}(k_0,\bm{k})\rho^{\beta\nu}_{f_2;db}(k_0,\bm{k})\right]\,.
\end{equation}
This quantity will vanish when the chemical potentials satisfy the relations
\begin{equation}
\mu_{f_1^a}-\mu_{f_1^c}=\mu_{f_2^d}-\mu_{f_2^b}.
\end{equation}
Then, demanding that each of the rates in \eqref{eq:formularates} vanish gives the conditions
\begin{eqnarray}
\mu_u-\mu_d &= & \mu_{\nu_\ell}-\mu_{\ell},\\
\mu_u-\mu_s &=& \mu_{\nu_\ell}-\mu_\ell,\\
\mu_u-\mu_s &=&\mu_u-\mu_d,\\ 
\mu_{\nu_{\ell_1}}-\mu_{\ell_1}&=&\mu_{\nu_{\ell_2}}-\mu_{\ell_2}.
\end{eqnarray}
Setting $\mu_{\nu_\ell}=0$, it is easy to check that this coincides with the beta equilibrium conditions in \eqref{eq:betaeq}. 

We will now expand \eqref{eq:gammacorr} to the lowest order around the equilibrium values, $\mu_{f^a}=\mu_{f^a}^{eq}+\delta\mu_{f^a}$. The thermal distributions have a pole at small frequencies, that we need to deal with. Let us introduce an arbitrary frequency $\omega_0$ such that $\delta\mu_{f_1^a}-\delta \mu_{f_1^c}\sim  \delta \mu_{f_2^d}-\delta \mu_{f_2^b}\ll \omega_0 \ll T  $. Then, there is a contribution that we can expand to linear order in the chemical potentials plus a correction from the poles
\begin{equation}\label{eq:linearmuexp0}
\gamma_{_{f_1^a f_2^b\to f_1^cf_2^d}}\approx \left(\delta \mu_{f_2^d}-\delta \mu_{f_2^b}-(\delta \mu_{f_1^a}-\delta \mu_{f_1^c})\right)\Lambda_{f_1^a f_2^b\to f_1^cf_2^d}^0+\delta \gamma_{_{f_1^a f_2^b\to f_1^cf_2^d}}\;,
\end{equation}
where, after defining $\Delta\mu^{eq}=\mu_{f_1^a}^{eq}- \mu_{f_1^c}^{eq}=\mu_{f_2^d}^{eq}- \mu_{f_2^v}^{eq}$, the coefficient $\Lambda$ is
\begin{equation}\label{eq:LambdaRate0}
\begin{split}
\Lambda_{_{f_1^a f_2^b\to f_1^cf_2^d}}^0=\,\,&\eta_{\mu\nu}\eta_{\alpha\beta}
  \int \frac{d^4 k}{(2\pi)^4}\rho^{\mu\alpha}_{f_1;ac}(k_0,\bm{k})\rho^{\beta\nu}_{f_2;db}(k_0,\bm{k}) \\
  &\times \left[\frac{1}{4T\sinh^2\left(\frac{k_0+\Delta \mu^{eq}}{2T} \right)}-\frac{T}{(k_0+\Delta\mu^{eq})^2}\Theta(\omega_0^2-(k_0+\Delta \mu^{eq})^2)\right].
  \end{split}
\end{equation}
We remind the reader that in this formula the fermion species are $f_1,f_2=q$ (quarks) or $l$ (leptons) and the possible flavors are $q^a=u,d,s$, $l^a=\nu_e,e,\nu_\mu,\mu$. For $k_0+\Delta \mu^{eq} \gtrsim T$ the function decays exponentially $\sim e^{-(k_0+\Delta \mu^{eq})/T}$, thus most of the contributions in the frequency integral will come from an interval of width $\sim T$ around $k_0+\Delta \mu^{eq}$. 

The correction to the rate originating from the poles is approximately
\begin{equation}
\begin{split}
\delta \gamma_{_{f_1^a f_2^b\to f_1^cf_2^d}}=&\eta_{\mu\nu}\eta_{\alpha\beta}
  \int_{|k_0+\Delta\mu^{eq}|<\omega_0} \frac{d^4 k}{(2\pi)^4} \rho^{eq\;\mu\alpha}_{f_1;ac}(k_0,\bm{k})\rho^{eq\;\beta\nu}_{f_2;db}(k_0,\bm{k}) \\
  &\times \left[\frac{T}{k_0+\Delta\mu^{eq}+\delta \mu_{f_1^a}-\delta \mu_{f_1^c}} -\frac{T}{k_0+\Delta\mu^{eq}+\delta \mu_{f_2^d}-\delta \mu_{f_2^b}}\right]\;.
  \end{split}
\end{equation}
We will use the principal value integral
\begin{equation}
\int_{-\omega_0-\Delta \mu^{eq}}^{\omega_0-\Delta \mu^{eq}} \frac{d k_0}{2\pi} \frac{T}{k_0+\Delta \mu^{eq}+\delta \mu} S(k_0) \approx \frac{T}{2\pi}S(-\Delta \mu^{eq})\log\frac{\omega_0+\delta\mu}{\omega_0-\delta \mu}\approx \frac{T}{\pi \omega_0}S(-\Delta \mu^{eq}) \delta \mu.
\end{equation}
Thus, we can write
\begin{equation}\label{eq:linearmuexp}
\gamma_{_{f_1^a f_2^b\to f_1^cf_2^d}}\approx \left(\delta \mu_{f_2^d}-\delta \mu_{f_2^b}-(\delta \mu_{f_1^a}-\delta \mu_{f_1^c})\right)\Lambda_{f_1^a f_2^b\to f_1^cf_2^d}\;,
\end{equation}
with
\begin{equation}\label{eq:LambdaRate}
\Lambda_{_{f_1^a f_2^b\to f_1^cf_2^d}}\approx \Lambda_{_{f_1^a f_2^b\to f_1^cf_2^d}}^0+\eta_{\mu\nu}\eta_{\alpha\beta}
 \frac{T}{\pi \omega_0} \int \frac{d^3 k}{(2\pi)^3}\rho^{eq\;\mu\alpha}_{f_1;ac}(-\Delta \mu^{eq},\bm{k})\rho^{eq\;\beta\nu}_{f_2;db}(-\Delta \mu^{eq},\bm{k}).
\end{equation}
If the spectral function vanishes at $k_0=-\Delta \mu^{eq}$ the last term in \eqref{eq:LambdaRate} is zero and one can take the $\omega_0\to 0$ limit, in which case the expression for \corr{the coefficient $\Lambda$} simplifies
\begin{equation}\label{eq:LambdaRateSimp}
\Lambda_{_{f_1^a f_2^b\to f_1^cf_2^d}}\approx \eta_{\mu\nu}\eta_{\alpha\beta}
  \int \frac{d^4 k}{(2\pi)^4} \frac{\rho^{eq\;\mu\alpha}_{f_1;ac}(-\Delta \mu^{eq},\bm{k})\rho^{eq\;\mu\alpha}_{f_2;db}(-\Delta \mu^{eq},\bm{k})}{4T\sinh^2\left(\frac{k_0+\Delta \mu^{eq}}{2T} \right)}.
\end{equation}

Combining the expression \eqref{eq:linearmuexp} with \eqref{eq:formularates} and \eqref{eq:ratecoefdef}, we arrive at the following formulas for the coefficients of the rates
\begin{subequations}\label{eq:Lambdacoefdef}
\begin{eqnarray}
\lambda_{ds} &\approx & 4 G_F^2 \sin^2\theta_C\cos^2\theta_C\; \Lambda_{ud\to su},\\
\lambda_{ud}^\ell &\approx & 4 G_F^2 \cos^2\theta_C\; \Lambda_{u\ell\to d\nu},\\
\lambda_{us}^\ell &\approx & 4 G_F^2 \sin^2\theta_C\; \Lambda_{u\ell\to s\nu},\\
\lambda_{e\mu} &\approx & 4 G_F^2\; \Lambda_{\mu\nu_e\to \nu_\mu e}.
\end{eqnarray}
\end{subequations}
When leptonic contributions can be neglected, as it is usually the case in the evaluation of the bulk viscosity in neutron stars, the only relevant coefficient is $\lambda_1=\lambda_{ds}$, following the conventions in \cite{Schmitt:2017efp}. In the remaining of the paper, we will give an estimate of $\lambda_1$ at strong coupling using a holographic model.

\section{Holographic model}\label{sec:holo}

We will use a holographic model dual to a $SU(N_c)$ gauge theory with $N_f$ flavors to compute the rate. We use a bottom-up approach, where we construct a gravitational action with the necessary ingredients to capture the physics of interest, but that does not correspond to a string theory construction where both sides of the duality are known.
The glue sector is given by an action consisting of just the Einstein-Hilbert action plus a negative cosmological constant
\be 
S_c = \frac{1}{16\pi G_5} \int d^5 x \, \sqrt{-g} \left[ R - 2\Lambda\right],\quad \Lambda=-\frac{6}{L^2}\;. 
\ee 
In the absence of matter, the maximally symmetric solution is an $AdS_5$ space with radius $L$. Including the flavor sector will modify the geometry, but it will still asymptote to the same $AdS_5$, so we can apply the usual holographic dictionary between the fields in the gravity theory and sources and operators in the gauge theory dual.

The flavor sector consists of $U(N_f)_L\times U(N_f)_R$ gauge fields, $L_M$ and $R_M$, dual to the flavor currents, and a complex scalar field $X$ in the bifundamental representation dual to a \corr{scalar} quark bilinear. The action for the flavor fields is the one commonly used for holographic QCD models, originating with \cite{Erlich:2005qh}
\corr{
\be \label{eq:Sf2}
S_f =\int d^5 x\,\sqrt{-g} \, {\rm Tr}\left[g_X^2\left(- \vert D X\vert^2 + \frac{3}{L^2} \vert X\vert^2\right) -\frac{1}{4 g_5^2} \left( F_{(R)}^2 + F_{(L)}^2\right) \right]\;.
\ee 
}
The field strength is defined as $F_{(A)MN} = \partial_M A_N - \partial_N A_M - i\left[ A_M,A_N\right] $. The covariant derivative of the scalar field is defined as follows 
\be 
D_N X = \partial_N X - i L_{N} X + i X R_{N} \,, \qquad D_N X^\dagger = \partial_N X^\dagger + i X^\dagger L_{N} - i  R_{N} X^\dagger\,.
\ee 
The mass of the scalar is fixed in such a way that the field is dual to a scalar operator of scaling dimensions $\Delta=3$, as corresponds to a quark bilinear.

The action is fixed up to an overall normalization of the scalar and gauge fields. In \cite{Erlich:2005qh,Cherman:2008eh} (see also \cite{Alvares:2011wb}) the normalization was fixed by matching the large Euclidean momentum form of current correlators derived from the holographic model to the OPE of QCD. 
Then, the matching to QCD in the UV fixes 
\corr{\begin{eqnarray}\label{eq:normfactors}
    \frac{1}{g_5^2}=\frac{N_c}{12\pi^2 L},\quad g_X^2L^3=\frac{N_c}{4\pi^2}\;.
\end{eqnarray}}

The equations of motion for the metric are Einstein's equations with a cosmological constant and the energy-momentum tensor sourced by the gauge fields and the scalars
\begin{equation}\label{eq:Einsteineq}
R_{MN}-\frac{1}{2}g_{MN}R+ g_{MN}\Lambda=\corr{8\pi G_5}T_{MN}.
\end{equation}
The equations of motion for the gauge and scalar fields read
\begin{subequations}
\bea
\frac{1}{\sqrt{-g}} D_M \left[\sqrt{-g} F^{MN}_{(R)}  \right] &=&  \corr{g_5^2 g_X^2}\, j^N_{(R)}  \,,\label{eq:EOMR}\\
\frac{1}{\sqrt{-g}} D_M \left[\sqrt{-g} F^{MN}_{(L)}  \right] &=& \corr{g_5^2 g_X^2}\, j^N_{(L)}\,,\label{eq:EOML}\\
\frac{1}{\sqrt{-g}} D_M \left[\sqrt{-g}  D^M X  \right] &=&   -\corr{\frac{3}{L^2}} \,X\,.\label{eq:eomX} 
\eea
\end{subequations}
where the bulk currents appearing in the equations of the gauge fields are
\begin{subequations}\label{eq:Xcurrents}
\bea
j^M_{(R)} & =& i \left[\left( D^M X\right)^\dagger X - X^\dagger D^M X\right]\,, \\
j^M_{(L)} & =& i \left[ \left( D^M X\right) X^\dagger - X \left(D^M X\right)^\dagger \right]\,.
\eea
\end{subequations}
We will look for solutions dual to a deconfined state of the gauge theory at nonzero temperature and baryon density, which are charged black brane geometries. In order to simplify the analysis we will turn off the scalar field, which amounts to having zero quark mass and condensate for all the flavors. In this case, imposing the condition of beta equilibrium fixes all quark chemical potentials to be equal, since deviations from symmetric quark matter due to nonzero quark masses are absent. Since we are ultimately interested in comparing with QCD results, we will fix the rank of the flavor fields $N_f=3$. 

\corr{When the scalars vanish, the charged black brane geometry is the preferred state of the system at large enough values of the chemical potential. There could also be other competing states where the scalar fields or off-diagonal components of the gauge fields acquire a non-trivial profile, corresponding to the spontaneous breaking of flavor symmetries. Since we are interested in comparing with results of unpaired quark matter we will neglect this possibility here.} 

\subsection{Background solution}

The metric of the charged black brane takes the general form, in Poincar\'e coordinates,
\begin{equation}\label{eq:blackbranemet}
ds^2 = \dfrac{L^2}{z^2}\left(-f(z)dt^2+\dfrac{dz^2}{f(z)}+d\Vec{x}\,^2\right), \qquad f(z)=1-M \frac{z^4}{z_h^4}+Q^2 \frac{z^6}{z_h^6}\;.
\end{equation}
The $AdS_5$ asymptotic boundary is at $z=0$, and the horizon is at the value where the black body factor vanishes $f(z_h)=0$. The values of $M$ and $Q$ depend on the background scalar and gauge fields. 

\corr{As we mentioned above, we will set all quark masses to zero, which allows to have a trivial solution of the scalar $X^{\rm bkg}=0$.} The background gauge fields dual to the quark densities are
\begin{equation}\label{eq:gaugefieldansatz}
(L_0^{\rm bkg})^a_{\ b}=(R_0^{\rm bkg})^a_{\ b}=\frac{\mu_q}{2}\left(1-\frac{z^2}{z_h^2}\right)\delta^a_{\ b}\;.
\end{equation}
Where $\mu_q=\mu_u=\mu_d=\mu_s$ is the quark chemical potential, which is the same for all flavors at beta equilibrium. \corr{Since in this case $\mu_e=0$, this implies that there is no net electron density, but the processes in \eqref{eq:weakreac} involving electron absorption or emission would still contribute, although we are neglecting them because they only give subleading corrections in the range of temperatures where the bulk viscosity is largest.} The backreaction over the metric is limited to the gauge fields, so the geometry is the AdS Reissner-Nordstrom charged black brane \eqref{eq:blackbranemet}, with 
\begin{equation}\label{eq:MQAdSRN}
M=1+Q^2\,,\qquad  Q^2=\frac{z_h^2\mu_q^2}{2}\;.
\end{equation}
The temperature in the dual field theory is equal to the Hawking temperature, given by
\begin{equation}\label{eq:TAdSRN}
    T = \dfrac{1}{\pi z_h}\left(1-\dfrac{Q^2}{2}\right)\;.
\end{equation}
One can use these formulas to obtain the position of the horizon as a function of the temperature and quark chemical potential
\corr{\begin{equation}\label{eq:zhAdSRN}
z_h=\frac{2}{\mu_q}\left(\sqrt{1+\left(\frac{\pi T}{\mu_q}\right)^2}-\frac{\pi T}{\mu_q} \right)\;.
\end{equation}}
Later we will often use
\begin{equation}
    \bar T= z_h T \approx \frac{2 T}{\mu_q}\;.
\end{equation}

\subsection{Two-point functions of flavor currents}

The spectral function entering the formulas for the rates can be obtained from the imaginary part of the retarded correlator of the left flavor currents. Following the usual holographic dictionary, we can extract the correlator from the solutions to the linearized equations of fluctuations. We will impose the boundary conditions 
\begin{equation}
\lim_{z\to 0} L_\mu(x,z)=l_\mu(x),\quad \lim_{z\to 0}R_\mu(x,z)=0,\quad \lim_{z\to 0} \frac{1}{z}X(x,z)=0.
\end{equation}
This turns off the sources for the right flavor currents and the quark bilinears and sets $l_\mu(x)$ as the source for the left flavor currents. The boundary asymptotic expansion of the Fourier transform of the left gauge fields is
\begin{equation}\label{eq:boundexpL}
(L_\mu)^a_{\ b}(\omega,\bm{k},z)\sim \left( \delta^{\ \nu}_\mu+(c_\mu^{\ \nu})_{ab} z^2 \log \frac{z}{L}+z^2 (g_\mu^{\ \nu})_{ab}+\cdots \right)(l_\nu)^a_{\ b}(\omega,\bm{k}).
\end{equation}
Where, \corr{having set the quark masses to zero,} 
\corr{\begin{equation}
    (c_\mu^{\ \nu})_{ab} = \dfrac{1}{2}\left(k_\alpha k^\alpha\delta_\mu^\nu - k_\mu k^\nu\right).
\end{equation}
}
The two-point function of the left flavor current can be obtained from the one-point function identifying the coefficient of the source $l_\mu$. The one-point function is extracted from the boundary value of the regularized canonical momentum
\corr{\begin{equation}
\vev{J_{q\,L}^\mu}=-\lim_{z\to 0} \frac{1}{ g_5^2}\left[\sqrt{-g} F_L^{z\mu}+\left(\log\frac{z}{L}+\log(L B)\right)\left(L D_\alpha\left(\sqrt{-\gamma} F_L^{\alpha \mu}\right)-g_X^2 g_5^2\sqrt{-\gamma} j_{(L)}^\alpha \right)\right],
\end{equation}}
where $\gamma_{\mu\nu}=g_{\mu\nu}$ is the induced metric on a radial slice and \corr{$B$} is an energy scale of the theory. The additional terms are obtained from the variation of the counterterm action
\corr{\begin{equation}
S_{c.t}=\left(\log\frac{z}{L}+\log(L B)\right)\int d^4 x\sqrt{-\gamma}\left[ \frac{L}{4g_5^2} F_L^{\mu\nu}F_{L\,\mu\nu}+g_X^2 L (D_\mu X)^\dagger D^\mu X\right]\;.
\end{equation}}
\corr{The energy scale $B$, that determines the coefficient of a finite counterterm, can be chosen arbitrarily. Its value is scheme dependent, depending on how the logarithmic divergence is cancelled. However, as shown in the formulas below it only enters in contact terms of the real part of the correlators, so it does not affect to the calculation of the spectral function. In addition to the terms shown here, one should also introduce a counterterm proportional to a quadratic potential for the scalar that is necessary to cancel further divergences when the quark masses are nonzero. In our case with massless quarks it can be safely ignored.}

Using the expansion \eqref{eq:boundexpL}, the one-point function to linear order in the fluctuation is
\corr{\begin{equation}
\vev{(J_{q\,L}^\mu)^a_{\ b}}= \frac{2L}{g_5^2} G^{\mu \nu}_{ab}(l_\nu)^a_{\ b},\qquad G^{\mu \nu}_{ab}=-\eta^{\mu\alpha}\left((g_\alpha^{\  \nu})_{ab}-(c_\alpha^{\ \nu})_{ab} \log\frac{L B}{e^{1/2}}\right)  \;.
\end{equation}}
Taking the variation with respect to the source gives the two-point function
\begin{equation}\label{eq:JLJL}
\vev{(J_{q\,L}^\mu)^a_{\ b} (J_{q\,L}^\nu)^{a'}_{\ b'} }= \frac{2L}{g_5^2} G_{ab}^{\mu \nu}\delta^a_{ b'}\delta^{a'}_b \;.
\end{equation}
\corr{The fit of the model to QCD at large Euclidean momentum \eqref{eq:normfactors} fixes the normalization constant  to the value}
\begin{equation}\label{eq:Normconst}
{\cal N}=\frac{2L}{g_5^2}=\frac{N_c}{6\pi^2}\;.
\end{equation}
The spectral function is obtained from the imaginary part of the retarded correlator, which can only receive contributions from the coefficient of the $O(z^2)$ term in the boundary expansion \eqref{eq:boundexpL}
\begin{equation}\label{eq:spectral}
\rho^{\mu\nu}_{ab,ba}=2{\cal N}\eta^{\mu\alpha}{\rm Im}\,(g_\alpha^{\  \nu})_{ab}\;.
\end{equation}
In order to compute the correlators of flavor currents involved in the formulas for the weak reaction rates, we must consider linearized fluctuations of the gauge fields around the background solution. We can restrict to fluctuations that are off-diagonal in flavor, and take into account that the background is diagonal. When the quark masses are zero and the scalar profile is trivial, the energy-momentum tensor of the gauge fields is at least quadratic in the fluctuations. Therefore, metric fluctuations can be turned off. Furthermore, the fluctuations of the left gauge fields decouple from the right gauge field and the scalar, so it is enough to restrict the analysis to the left gauge fields.

\subsection{Linearized fluctuations}

\corr{We will work in the radial gauge $L_z=0$ and expand the fluctuations in Fourier modes
\begin{equation}
L_\mu(x,z)=\int \frac{d\omega}{2\pi}\int \frac{d^3 k}{(2\pi)^3} e^{-i\omega x^0+i\bm{k}\cdot\bm{x}} L_\mu(\omega,\bm{k},z)\;.
\end{equation}
Our fluctuations correspond to the off-diagonal flavor components $L_\mu\equiv (L_\mu)^a_{\ b}$, but we will omit flavor indices unless there is some ambiguity.}

It will be convenient to rescale coordinates by the horizon position 
\begin{equation}
z=z_h Z,\quad  x^\mu=z_h X^\mu\;,
\end{equation}
and work with the dimensionless frequency and  momentum  
\begin{equation}
\bar\omega=z_h \omega,\quad \bm{\bar k}=z_h \bm{k}\;,  (\bar k=|\bm{\bar{k}}|)\;.
\end{equation}
\corr{We will also define a dimensionless temperature $\bar T=z_h T$. The equations of motion for the linearized fluctuations of the gauge fields in these coordinates are collected in Appendix~\ref{app:eoms}.}

We are interested in temperatures much smaller than the chemical potential $T/\mu_q \ll 1$. Since the integral over frequencies that determines the rates \eqref{eq:LambdaRate} is concentrated on frequencies smaller or of the order of the temperature, we can use a small frequency approximation $\omega/\mu_q \ll 1$. 

We will use the approach of \cite{Faulkner:2009wj} to compute low-frequency correlators. When the temperature is zero, the background solution is an extremal black brane that develops an $AdS_2$ throat at the horizon. One can then divide the geometry into an inner and outer region relative to this throat. When the temperature is non-zero, but very small compared to the chemical potential, there is still a throat but the inner geometry is a non-extremal $AdS_2$ with a horizon. From \eqref{eq:TAdSRN}, we see that the near-extremal solutions correspond to $Q^2\lesssim 2$. Let us parametrize
\begin{equation}\label{eq:zetacoord}
Q^2=\frac{2}{Z_*^6},\qquad Z_*-1=|\bar \omega| \frac{Z_*^2}{12 \zeta_0},\qquad Z_*-Z=|\bar \omega| \frac{Z_*^2}{12 \zeta}\,.
\end{equation}
The temperature in units of the horizon position is $\pi \bar T=1-Z_*^{-6}$. This gives the approximate relation
\begin{equation}
\zeta_0\approx \frac{|\bar\omega|}{2\pi \bar T}=\frac{|\omega|}{2\pi T}\;.
\end{equation}
Expanding the metric \eqref{eq:blackbranemet} for $|\bar\omega|\to 0$ and $Z_*\approx 1$ (since $\bar T\ll 1$), one gets
\begin{equation}\label{eq:AdS2metric}
ds^2\approx \frac{L^2}{12 \zeta^2}\left(-F(\zeta)d\bar\tau ^2+\frac{d\zeta^2}{F(\zeta)} \right)+L^2d\Vec{X}\,^2\,, \qquad F(\zeta)=1-\frac{\zeta^2}{\zeta_0^2},\quad \bar\tau=|\bar \omega| X^0=|\omega| x^0 \;.
\end{equation}
This is $AdS_2\times \mathbb{R}^3$ Schwarzschild geometry with an $AdS_2$ radius $L^2_{_{AdS_2}}=L^2/12$ and a horizon at $\zeta=\zeta_0$. In the zero temperature limit $Z_*\to 1$, $\zeta_0\to \infty$ the throat extends to values $\zeta\to \infty$.

\corr{The equations for linearized fluctuations can be solved in the inner region, $L_\mu^{\rm in}$, imposing an ingoing boundary condition at the horizon $\zeta\to \zeta_0$,  and in the outer region, $L_\mu^{\rm out}$, taking the zero temperature limit where the geometry becomes the extremal $AdS_5$ Reissner-Nordstrom, \eqref{eq:blackbranemet} with $Q=\sqrt{2}$. The calculation of the solutions and connection coefficients up to $O(\omega^2/\mu_q^2)$ corrections is detailed in Appendix~\ref{app:solrn}, we summarize here the main results.}  

In the region towards the end of the $AdS_2$ throat, \corr{$\zeta\to 0$}, the solutions have an expansion
\begin{equation}
\corr{L_\mu^{\rm in} }\sim A_\mu^-\zeta^{\Delta_-}+A_\mu^+ \zeta^{\Delta_+}\,,\qquad \Delta_+>\Delta_-.
\end{equation} 
This matches with the behaviour of solutions in the outer region close to the horizon $Z\to 1$
\begin{equation}
\corr{L_\mu^{\rm out}} = C^-_\mu \eta_-(Z)+C^+_\mu \eta_+(Z),\quad \eta_\pm(Z)\sim (1-Z)^{-\Delta_\pm}\;,
\end{equation}
which allows the identification
\begin{equation}
C^\pm_\mu=A_\mu^\pm \left(\frac{|\bar\omega|}{12}\right)^{\Delta_\pm}\;.
\end{equation}
The ingoing condition fixes a relation between $A_\mu^+$ and $A_\mu^-$ that translates into a relation between $C_\mu^+$ and $C_\mu^-$:
\begin{equation}
C_\mu^+=\corr{\mathcal{G}_L(\omega,k)} C_\nu^-\;,
\end{equation}
with \corr{$\mathcal{G}_L(\omega,k)$} usually interpreted as an infrared (IR) Green's function associated with a $0+1$ CFT dual to the $AdS_2$ gravitational theory. This IR function determines the Green's function of the full theory. In order to see this, we need to translate the above relation between coefficients to the outer normalizable and non-normalizable solutions
\begin{equation}
\corr{L_\mu^{\rm out}} = l_\mu \phi_{NN}(Z)+g_\mu \phi_N(Z),
\end{equation}
with expansions close to the boundary $Z\to 0$
\begin{equation}
\phi_{NN}(Z)\sim 1+\frac{\bar k^2}{2}Z^2\log Z+O(Z^4),\ \ \phi_N(Z)=Z^2+O(Z^4)\;.
\end{equation}
The two sets of outer solutions, \corr{$\{\eta_+,\eta_-\}$ and $\{\phi_{NN},\phi_N\}$}, are related by the connection coefficients
\begin{equation}
\eta_\pm(Z)=\alpha_\pm \phi_{NN}(Z)+\beta_\pm \phi_N(Z)\;. 
\end{equation}
Thus, the full solution is
\begin{equation}
\corr{L_\mu^{\rm out}} =C_\mu^-\left( \eta_-(Z)+\corr{\mathcal{G}_L}\eta_+(Z)\right)=C_\mu^-\left(\alpha_- +\alpha_+\corr{\mathcal{G}_L}\right)\phi_{NN}(Z)+C_\mu^-\left(\beta_- +\beta_+\corr{\mathcal{G}_L}\right)\phi_N(Z)\;.
\end{equation}
Fixing the coefficient of the non-normalizable solution to be the source $l_\mu$, we get that the coefficient of the normalizable term is
\begin{equation}
g_\mu=\frac{\beta_- +\beta_+\corr{\mathcal{G}_L}}{\alpha_- +\alpha_+\corr{\mathcal{G}_L}}\,l_\mu\;.
\end{equation} 
More precisely, the connection coefficients and IR Green's function depend on whether the fluctuation is transverse or longitudinal. It turns out that at leading order in $\omega/T$ the longitudinal components do not contribute.

\subsection{Spectral function}

Since only the transverse components of the left gauge fields are relevant at low frequencies, the spectral function determined by \eqref{eq:spectral} is ($\rho^{00}\approx 0$, $\rho^{0i}\approx 0$),
\begin{equation}\label{eq:spectralfunc}
\rho^{ij}_{ab,ba}\approx {\cal N}\frac{\mu_q^2}{2}\left(\delta^{ij}-\frac{k^i k^j}{k^2}\right)\rho_L\left(\frac{\omega}{2\pi T},\bar k,\bar T\right)\,,\quad \rho_L\left(\frac{\omega}{2\pi T},\bar k,\bar T\right)=\dfrac{\beta_+\alpha_- - \beta_-\alpha_+}{\left(\alpha_-+ \alpha_+\text{Re}\,\mathcal{G}_L\right)^2 + \left(\alpha_+\right)^2\left(\text{Im}\,\mathcal{G}_L\right)^2}\text{Im}\,\mathcal{G}_L\;.
\end{equation}
The IR Green's function of the transverse modes of the left gauge fields is, \corr{up to corrections suppressed by factors of $T/\mu_q$,}
\begin{equation}\label{eq:GL}
\mathcal{G}_L(\bar\omega,\bar k,\bar T)\approx \frac{\Gamma \left(-\nu_k\right) \Gamma \left(\frac{1}{2}-i \,\frac{\omega}{2\pi T}+\nu_k\right)}{\Gamma \left(\nu_k\right) \Gamma \left(\frac{1}{2}-i \,\frac{ \omega}{2\pi T}-\nu_k\right)}\left(\frac{\pi \bar T}{12} \right)^{2\nu_k}.
\end{equation}
\corr{Where we have introduced $\nu_k$ for convenience, defined as
\begin{equation}
\nu_k=\sqrt{\frac{1}{4}+\frac{\bar{k}^2}{12}}\;.
\end{equation}
}
At large spatial momentum the IR Greens' function approaches
\begin{equation}\label{eq:GLlargek}
\mathcal{G}_L(\bar\omega,\bar k,\bar T)\sim -\frac{\cos\left(\frac{\pi \bar k}{\sqrt{12}}+i\frac{\omega}{2T}\right)}{\sin\left(\frac{\pi \bar k}{\sqrt{12}}\right)}\left(\frac{\pi \bar T}{12} \right)^{\frac{\bar k}{\sqrt{3}}}\;. 
\end{equation}
At low temperatures, there is a strong suppression due to the $\bar T$-dependent factor. When $T/\mu_q\to 0$ ($\bar T\to 0$), there is a strong suppression of the correlator already at small momenta
\begin{equation}
\mathcal{G}_L\sim \left(\frac{\pi \bar T}{12} \right)e^{-\frac{\bar k^2}{12}\log\frac{12}{\pi \bar T}}\,.
\end{equation}
The support of the Green's function is concentrated on values $\bar k\lesssim |\log(T/\mu_q)|^{-1} \ll 1$. In this case, we can approximate the Green's function by
\begin{equation}\label{eq:GLlowT}
\mathcal{G}_L(\bar\omega,\bar k,\bar T)\approx \left(\frac{\pi \bar T}{12} \right)e^{-(2\nu_k-1)\log\frac{12}{\pi \bar T}}\left(i\frac{\omega}{\pi T}+ (2\nu_k-1)\right)+O((2\nu_k-1)^2,i\bar \omega (2\nu_k-1))\;. 
\end{equation}
The connection coefficients are real and only depend on the momentum. We have computed them numerically and determined their asymptotic behaviour at large spatial momentum (c.f. Appendix~\ref{app:solrn})
\begin{align}\label{eq:WKBalphabetaMT}
    &\alpha_\pm \sim \mp 6^{\mp\frac{\bar{k}}{\sqrt{12}}}\bar{k}^{-1/2} ,\\
    &\beta_\pm \sim \mp 6^{\mp\frac{\bar{k}}{\sqrt{12}}}\,\bar{k}^{3/2}\,\log\bar{k}.
\end{align}
Considering the large momentum asymptotic behaviour of the IR Green's function \eqref{eq:GLlargek} and connection coefficients \eqref{eq:WKBalphabetaMT}, we see that the spectral function vanishes exponentially fast at large spatial momentum. A special point is $\bar k=3$ ($\nu_k=1$), where $\mathcal{G}_L$ diverges and the spectral function has its first zero. For a numerical estimate, one can thus restrict the momentum integral to smaller values.

When $T/\mu_q\to 0$ we can approximate the connection coefficients by their $\bar k\to 0$ values, that we can compute analytically
\begin{align}\label{eq:LowTalphabetaMT}
    &\alpha_- \approx 1+O\left(\nu_k-\frac{1}{2}\right), \quad \beta_-\approx -\frac{3}{2}(2\nu_k-1)+O\left[\left(\nu_k-\frac{1}{2}\right)^2\right],\\
    &\alpha_+\approx  \frac{4}{3(2\nu_k-1)}+O(1),\quad \beta_+\approx 4+O\left(\nu_k-\frac{1}{2}\right).
\end{align}
Thus $\beta_+\alpha_--\beta_-\alpha_+\approx 6$ and
\begin{equation}\label{eq:LowTrho}
\rho_L\left(\frac{\omega}{2\pi T},\bar k,\bar{T}\right)\approx \dfrac{6(2\nu_k-1)^2}{\left(2\nu_k-1\right)^2 + \left(\frac{4 A_k(\bar T) \omega}{3\pi T}\right)^2}  A_k(\bar T)\frac{\omega}{\pi T}\approx 6A_k(\bar T)\frac{\omega}{\pi T}\;.
\end{equation}
Where we have defined
\begin{equation}
A_k(\bar T)=\left(\frac{\pi \bar T}{12} \right)e^{-(2\nu_k-1)\log\frac{12}{\pi \bar T}}\;.
\end{equation}

\begin{figure}
\center
\includegraphics[width = 0.5\textwidth]{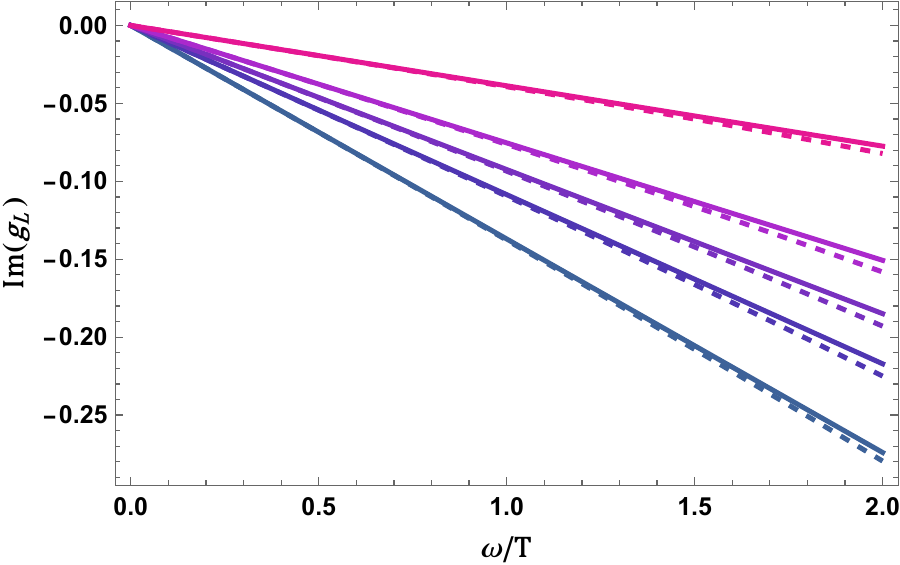}\caption{Plot of the imaginary part of the full  $g_L$ function (dashed) and its leading-order approximation in $\omega/T$ (solid) for $\nu_k =0.6,0.7,0.75,0.8,0.9$ which correspond to colors from blue to light purple.}
\label{fig:ImIRG}
\end{figure}

\begin{figure}
    \centering
    \begin{subfigure}{0.45\textwidth}
        \centering
        \includegraphics[width=\textwidth]{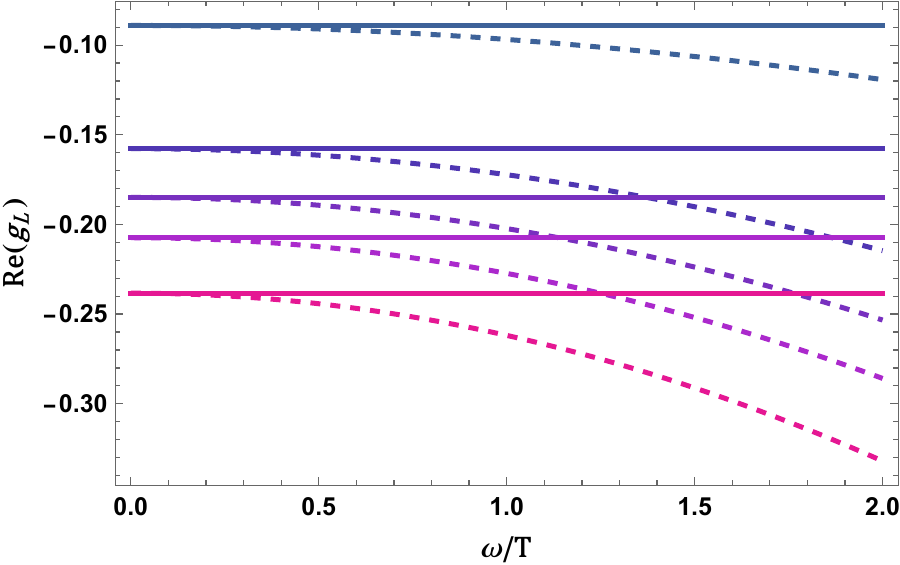}
        \caption{Plot of the real part of the full $g_L$ function (dashed) and its leading order approximation in $\omega/T$ (solid) for $\nu_k =0.6,0.7,0.75,0.8,0.9$ which correspond to colors from blue to light purple.}
        \label{fig:ReIRG0}
    \end{subfigure}
    \hfill
    \begin{subfigure}{0.45\textwidth}
        \centering
        \includegraphics[width=\textwidth]{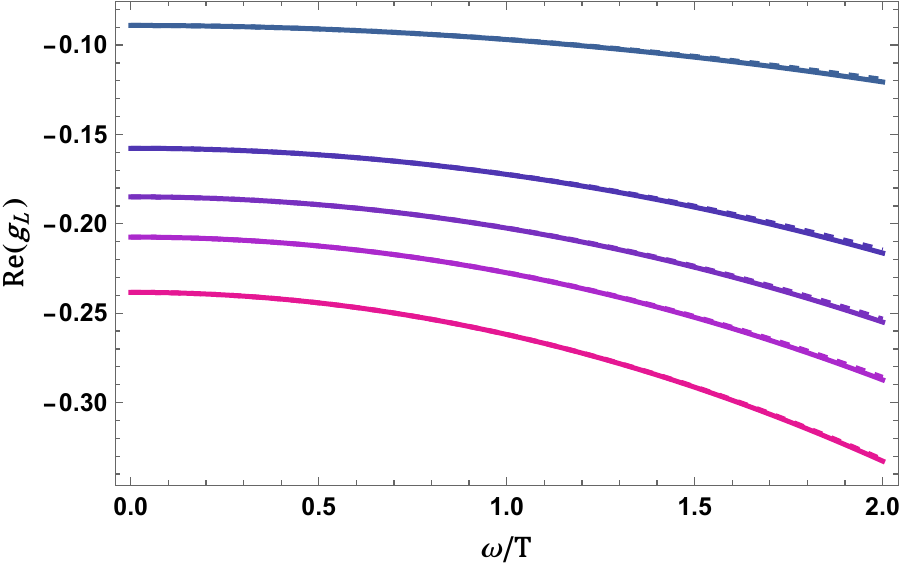}
        \caption{Plot of the real part of the full $g_L$ function (dashed) and its second-order approximation in $\omega/T$ (solid) for $\nu_k =0.6,0.7,0.75,0.8,0.9$ (colors from blue to light purple).}
        \label{fig:ReIRG2}
    \end{subfigure}
    \caption{Plots of the real part of the full $g_L$ (dashed lines) and its approximated expressions up to second-order in $\omega/T$ (solid lines).}
\end{figure}

Numerically, at low frequencies the IR Green's function is well approximated by
\begin{align}\label{eq:lowT2}
\mathcal{G}_L(\bar\omega,\bar k,\bar T)\approx &\frac{\Gamma \left(-\nu_k\right) \Gamma \left(\frac{1}{2}+\nu_k\right)}{\Gamma \left(\nu_k\right) \Gamma \left(\frac{1}{2}-\nu_k\right)}\left(\frac{\pi \bar T}{12} \right)^{2\nu_k}\bigg(1-\frac{i}{2}\tan(\pi \nu_k)\frac{\omega}{T} +\\ \nonumber
&+ \dfrac{ \pi ^2 \tan ^2(\pi  \nu_k )-\psi ^{(1)}\left(\frac{1}{2}-\nu_k \right)+\psi ^{(1)}\left(\frac{1}{2}+\nu_k\right)}{8\pi^2}\frac{\omega^2}{T^2}\bigg)+O\left(\frac{\omega^3}{T^3}\right),
\end{align}
where $\psi^{(1)}$ is a polygamma function. In order to see how good is the approximation we plot the real and the imaginary part of the full IR Green function and its approximated expression. We define a reduced IR Green function
\begin{equation}
    g_L(\bar\omega,\bar k) = \frac{\Gamma \left(\nu_k\right)}{\Gamma \left(-\nu_k\right) }\left(\frac{12}{\pi \bar T} \right)^{2\nu_k}\mathcal{G}_L(\bar\omega,\bar k,\bar T),
\end{equation}
so that the plots will be independent of $\bar{T}$. In Figure \ref{fig:ImIRG} we plot the full imaginary part of $g_L$ (dashed lines) with the approximated formula (solid lines) for some values of $\nu_k$, as a function of $\omega/T \in (0,2)$. Then, we plot the full real part of $g_L$ (dashed lines) with the approximated formula (solid lines), as a function of $\omega/T \in (0,2)$, keeping just the first term in the expansion, Figure \ref{fig:ReIRG0}, and then adding also the quadratic correction, Figure \ref{fig:ReIRG2}. The values of $\nu_k$ are $0.6,0.7,0.75,0.8,0.9$ which correspond to colors from blue to light purple. We observe that as $k$ becomes smaller, the real part becomes better approximated by the leading term.

\section{Estimate of weak reaction rates at strong coupling}\label{sec:rate}

\begin{figure}
\center
\includegraphics[width = 0.6\textwidth]{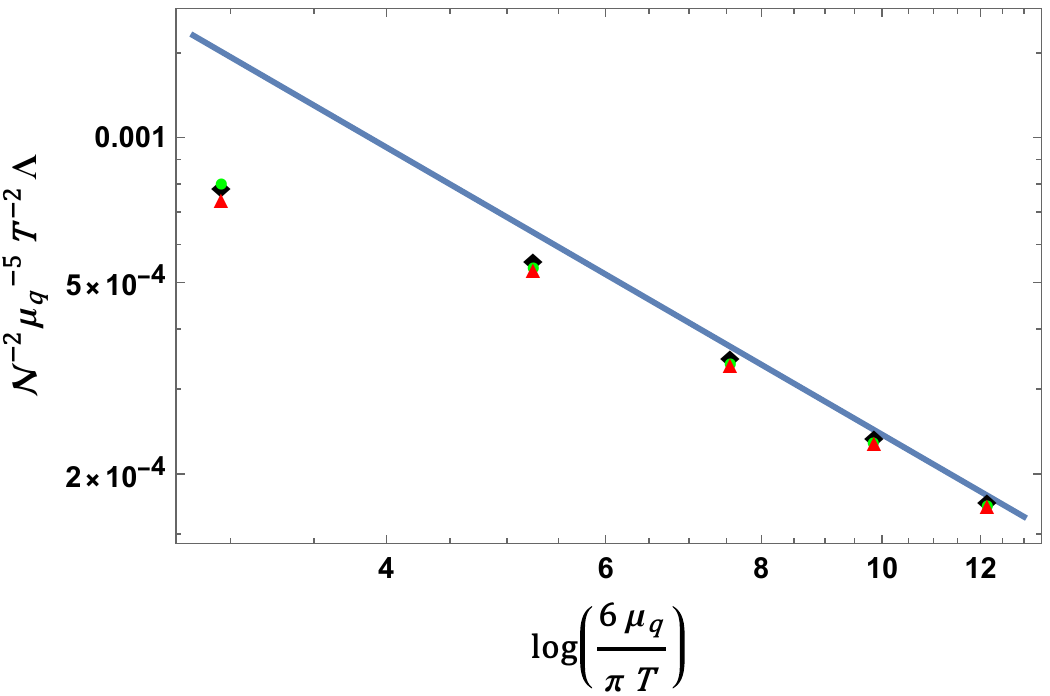}\caption{Log-log plot of the weak reaction rate multiplied by $\mathcal{N}^{-2}\mu_q^{-5}T^{-2}$ as function of $\log\left(\frac{6\mu_q}{\pi T}\right)$. The solid line described the approximated analytic results of Equation \eqref{eq:rateanalytic}. The symbols are the results of the numerical integration of the formula \eqref{eq:ratefullformula} for $T/\mu_q=10^{-1},10^{-2},10^{-3},10^{-4}$, taking the leading order approximation of the IR Green function (green dots), the second order approximation (red triangles) and the full result (black diamonds).}
\label{fig:Rate}
\end{figure}

We will proceed to estimate the weak rate. Since all the quark chemical potentials are equal at beta equilibrium and the spectral function \eqref{eq:spectralfunc} vanishes at zero frequency, we can use the simplified formula \eqref{eq:LambdaRateSimp} to compute the rate. Taking into account that only the transverse components contribute to leading order, and rescaling the variables in the momentum integral \corr{$\omega=k_0=2 \pi T s$}, $k=\mu_q \bar k/2$, 
\begin{equation}\label{eq:ratefullformula}
\Lambda_{ud\to us}\approx {\cal N}^2\frac{\mu_q^7}{2^7\pi^2}
  \int_{-\infty}^\infty ds \int_0^\infty d\bar k\, \bar k^2\, \frac{\rho_L^2(s,\bar k,\bar T)}{\sinh^2(\pi s)}\,.
\end{equation}
Introducing \eqref{eq:LowTrho}, and changing variables to $\bar k^2=3(4\nu^2-1)$,
\begin{equation}
\Lambda_{ud\to us}\approx {\cal N}^2\frac{6^2\mu_q^7}{2^7\pi^2}2\left(\frac{\pi \bar T}{12} \right)^2
  \int_0^\infty ds \frac{(2s)^2}{\sinh^2(\pi s)} \int_{1/2}^\infty d\nu\,12 \nu \sqrt{3(4\nu^2-1)}\, e^{-2(2\nu-1)\log\frac{12}{\pi \bar T}}\,.
\end{equation}
\begin{equation}
\Lambda_{ud\to us}\approx {\cal N}^2\frac{3\sqrt{3}\mu_q^5 T^2}{4}
  \int_0^\infty ds \frac{s^2}{\sinh^2(\pi s)} \int_{1/2}^\infty d\nu\, \nu \sqrt{(4\nu^2-1)}\, e^{-2(2\nu-1)\log\frac{12}{\pi \bar T}}\,.
\end{equation}
We have
\begin{equation}
\int_0^\infty ds \frac{s^2}{\sinh^2(\pi s)}=\frac{1}{6\pi}
\end{equation}
Changing variables $\nu=\frac{1}{2}+u$,
\begin{equation}
\Lambda_{ud\to us}\approx {\cal N}^2\frac{\sqrt{3}\mu_q^5 T^2}{8\pi}
   \int_0^\infty du\, (1+2u) \sqrt{u(u+1)}\, e^{-4u\log\frac{12}{\pi \bar T}}\,.
\end{equation}
We use that the main contributions to the integral come from $u\ll 1$:
\begin{equation}
   \int_0^\infty du\, (1+2u) \sqrt{u(u+1)}\, e^{-4u\log\frac{12}{\pi \bar T}}\approx \int_0^\infty du\, u^{1/2} \, e^{-4u\log\frac{12}{\pi \bar T}}=\frac{\Gamma(3/2)}{\left(4\log\frac{12}{\pi \bar T}\right)^{3/2}}=\frac{\sqrt{\pi}}{2\left(4\log\frac{12}{\pi \bar T}\right)^{3/2}}\,.
\end{equation}
Then,
\begin{equation}\label{eq:rateanalytic}
\Lambda_{ud\to us}\approx \frac{{\cal N}^2}{128}\sqrt{\frac{3}{\pi}}\mu_q^5 T^2\left(\log\frac{6\mu_q}{\pi T}\right)^{-3/2}
   \,.
\end{equation}
We have computed numerically the rate for values of the temperature $T/\mu_q=10^{-1},10^{-2},10^{-3},10^{-4}$, considering the full dependence of the connection coefficients, and several approximations for the IR Green's function, namely the leading order \eqref{eq:GLlowT}, the second order approximation \eqref{eq:lowT2}, and the full result \eqref{eq:GL}. In the numerical integration over momentum, we cut the integral at $\nu_k=1$, where the spectral function has its first zero. The exponential suppression already present at this value of the spatial momentum makes higher momentum contributions to be exceedingly small, so we believe this is a good approximation. We show our numerical results together with the analytic approximation at low temperatures \eqref{eq:rateanalytic} in Figure~\ref{fig:Rate}. We see that indeed the analytic formula approximates well the numerical result at low temperatures, even up to $T/\mu_q \lesssim 10^{-2}$, but for $T/\mu_q\sim 10^{-1}$ the deviations are already sizeable, although the analytic formula still serves as an estimate of the order of magnitude.

\section{Discussion and outlook}\label{sec:conc}

In this work, we derived the formulas \eqref{eq:formularates} and \eqref{eq:gamma} for the weak equilibration rate of quark flavor densities $n_a$ ($a=u,d,s$). \corr{These formulas describe the equilibration rates near thermal equilibrium in terms of the spectral function of left-handed currents, to leading order in the Fermi coupling. The result, derived purely through field theory techniques, stems from the modification of Ward identities for flavor currents by symmetry-breaking terms induced by the weak interaction. Up to possible corrections by effective eight-fermion interactions, it is exact in the QCD coupling, with all non-perturbative dynamics of the strongly coupled theory encapsulated in the spectral function.} Although we have focused on quark matter, the formulas are derived in an explicitly gauge-invariant way and rely solely on symmetry arguments. Thus, they should be valid for any state of QCD at nonzero temperature, including hadronic (confined) phases. In a confined phase, the natural degrees of freedom are baryons rather than quarks. Baryons are composed of three quarks and can be grouped according to their flavor composition. Therefore, the chemical potential for a baryon $B^{abc} \sim q^a q^b q^c$ would be $\mu_{B^{abc}} = \mu_a + \mu_b + \mu_c$, with $\mu_{a,b,c}$ being the quark chemical potentials corresponding to each flavor. The associated baryon densities should satisfy the conditions
\begin{equation}
n_a = 3 n_{B^{aaa}} + 2 \sum_{b \neq a} n_{B^{aab}} + \sum_{b,c \neq a} n_{B^{abc}},
\end{equation}
where the order of the flavor labels is immaterial, and $n_{B^{abc}}$ would, in principle, include all contributions from baryon resonances with the same flavor composition and with a mass below the value of the baryon chemical potential, assuming the decay channels to lower resonances become closed. With these identifications, one can combine the expressions in \eqref{eq:formularates} to obtain the rates for the evolution of the density of different baryon species, or rather for the evolution of some combinations of them.

Returning to our analysis, we have \corr{used the general formulas for the rates in a particular model of unpaired quark matter} \corr{with a holographic dual description. Using the holographic dual, we have obtained the leading order low-temperature behavior of the weak reaction rate in a large-density state of strongly coupled matter at beta equilibrium. } For simplicity, we have only considered massless quarks, from which we were able to extract an analytic expression \eqref{eq:rateanalytic} for the rate coefficient as defined in \eqref{eq:ratecoefdef} and \eqref{eq:Lambdacoefdef}, resulting in
\begin{equation}\label{eq:lambdadsresult}
\lambda_{ds} \approx G_F^2 \sin^2 \theta_C \cos^2 \theta_C \frac{{\cal N}^2}{32} \sqrt{\frac{3}{\pi}} \mu_q^5 T^2 \left( \log \frac{6 \mu_q}{\pi T} \right)^{-3/2}.
\end{equation}
A fit to the large momentum behavior of QCD correlators fixes the normalization constant to ${\cal N} = N_c / (6 \pi^2)$, Eq. \eqref{eq:Normconst}. We can compare this with the perturbative result \cite{Heiselberg:1986pg,Heiselberg:1992bd,Madsen:1993xx} including the leading QCD correction \cite{Schwenzer:2012ga}
\begin{equation}
\lambda_{ds} \approx G_F^2 \sin^2 \theta_C \cos^2 \theta_C \frac{64}{4 \pi^2} \mu_q^5 T^2 \left(1 + \frac{4 \alpha_s}{9 \pi} \log \frac{\kappa \, \alpha_s \mu_q}{T} \right)^4,
\end{equation}
where $\kappa \approx  0.28(2/3\pi)^{1/2} \approx 0.13$. The logarithmic enhancement of the rate at low temperatures originates from a non-Fermi-liquid behavior of quark matter due to long-range interactions (see also \cite{Gerhold:2004tb,Schafer:2004zf}). Although the logarithm should not be too large in order for the perturbative expansion to be valid, the fourth power with which it appears can produce a significant effect on the rate. Considering the strong coupling result, the holographic dual geometry has also been conjectured to describe a non-Fermi (``marginal'') liquid \cite{Faulkner:2009wj}. However, the logarithmic factor operates in the opposite way, suppressing the rate. Furthermore, compared to the perturbative result, the logarithm will be much larger, as there are no coupling-dependent factors in the argument suppressing the ratio $\mu_q/T$. As shown in Figure~\ref{fig:Rate}, the analytic formula \eqref{eq:lambdadsresult} provides a good approximation for temperatures $T/\mu_q$ below $10^{-2}$, but at higher temperatures, corrections to this formula start to become significant. However, their effect is to further suppress the rate.

Several improvements could be applied to the model, with perhaps the most significant being the inclusion of non-zero quark masses. In the perturbative calculation of the rate, the mass enters the dispersion relation of the fermions in the scattering amplitude and removes the low-energy contributions from the massive quarks. In the holographic model, the scalar field dual to the quark bilinears would be non-zero, and its backreaction can change the near-horizon geometry, thus affecting the rate behavior. A generic expectation is that the $AdS_2$ throat in the near-horizon geometry would be replaced by a hyperscaling Lifshitz geometry (see \cite{Taylor:2015glc} and references therein), which would probably result in the rate becoming proportional to a different power of the temperature depending on the dynamical exponent and the hyperscaling parameters.

Another extension of the model would be to study phases of paired quark matter, as quark pairing can affect transport properties \cite{Sad:2006egl,Alford:2006gy,Alford:2008pb}. A paired state would have a quark condensate with a non-trivial flavor structure, which could be captured by turning on a profile of the scalar field reproducing the same structure.

In conclusion, our study provides a framework for understanding weak interaction rates in strongly coupled quark matter and highlights the importance of non-perturbative methods in studying dense QCD phases. Future work should focus on refining these models, incorporating more realistic features such as non-zero quark masses and quark pairing, to enhance our understanding of neutron star interiors and their observational signatures.

\section*{Acknowledgments}

We thank Niko Jokela for useful discussions. This work is partially supported by the AEI and the MCIU through the
Spanish grant PID2021-123021NB-I00. A.O. thanks the University of Oviedo for the kind hospitality during the preparation of this work. 

\appendix

\section{Non-conservation equations of flavor currents}\label{app:flavorbreak}

With three massless flavors of quarks $q^i=(u,d,s)$ and one flavor of massless leptons $l^a=(\nu_e,e)$ (with only left-handed component for the neutrino), the flavor symmetry group is $SU(3)_{q_L}\times SU(3)_{q_R}\times SU(2)_{l_L}\times U(1)_{e_R}$. Unequal mass terms for the quarks and the electron break explicitly the flavor group to $U(1)^5$, where the quarks and electron $U(1)$s are vector-like.

The EW sector, in particular, W-boson exchange introduces a term in the effective action of the form (Fermi model)
\begin{equation}\label{eq:FermiActionAPP}
\Delta {\cal L}_{EW}=-2\sqrt{2} G_F (J_{ch}^\mu)^\dagger J_{ch\,\mu},
\end{equation}
where the $J_{ch}^\mu$ is the left-charged current
\begin{equation}
J_{ch}^\mu=\overline{\nu}_{e\,L} \gamma^\mu e_L+\cos \theta_C\, \overline{u}_L\gamma^\mu d_L+\sin \theta_C\, \overline{u}_L\gamma^\mu s_L,\ \
\end{equation}
with $\theta_C$ the Cabbibo angle, which gives the change of basis between mass and flavor eigenstates. Let us introduce the notation
\begin{equation}
(J_q^\mu)^i_{\ j}=\bar{q}_j \gamma^\mu q^i,\ \ (J_l^\mu)^a_{\ b}=\bar{l}_b \gamma^\mu l^a\,.
\end{equation}
Then, in terms of current components,
\begin{equation}
J_{ch}^\mu=(J_{l\,L}^\mu)^{2}_{\ 1}+\cos \theta_C (J_{q\,L}^\mu)^{2}_{\ 1}+\sin \theta_C  (J_{q\,L}^\mu)^{3}_{\ 1},\ \
\end{equation}
Where $J_{l\,L}^\mu$ is the current for the $SU(2)_L$ lepton symmetry and $J_{q\,L}^\mu$ is the current for the $SU(3)_L$ flavor symmetry. The EW term becomes
\begin{equation}
\begin{split}
\Delta {\cal L}_{EW}=&-2\sqrt{2} G_F \left[ \cos^2\theta_C (J_{q\,L}^\mu)^{ 2}_{\ 1} (J_{q\,L\,\mu})^{1}_{\ 2}+\sin^2 \theta_C  (J_{q\,L}^\mu)^{ 1}_{\ 3} (J_{q\,L\,\mu})^{3}_{\ 1}\right.\\
&+\cos\theta_C \sin\theta_C\left( (J_{q\,L}^\mu)^{ 2}_{\ 1} (J_{q\,L}^\mu)^{ 1}_{\ 3}+(J_{q\,L}^\mu)^{3}_{\ 1}(J_{q\,L}^\mu)^{ 1}_{\ 2}\right)\\ 
&+\cos\theta_C \left((J_{l\,L}^\mu)^{2}_{\ 1}(J_{q\,L\,\mu})^{1}_{\ 2}+(J_{l\,L}^\mu)^{1}_{\ 2}(J_{q\,L\,\mu})^{2}_{\ 1}\right)\\
&\left.+\sin\theta_C \left((J_{l\,L}^\mu)^{2}_{\ 1}(J_{q\,L\,\mu})^{ 1}_{\ 3}+(J_{l\,L}^\mu)^{ 1}_{\ 2}(J_{q\,L\,\mu})^{3}_{\ 1}\right)\right]\,.
\end{split}
\end{equation}
Let us neglect the terms coupling quarks and leptons. We can rewrite the remaining terms in the following way
\begin{equation}
\begin{split}
\Delta {\cal L}_{EW}=&-2\sqrt{2} G_F \left[ \cos^2\theta_C \,\text{tr}\left(P_{11} J_{q\,L}^\mu P_{22} J_{q\,L\,\mu}  \right) +\sin^2 \theta_C \, \text{tr}\left(P_{11} J_{q\,L}^\mu P_{33} J_{q\,L\,\mu}  \right)\right.\\
&\left.+\cos\theta_C \sin\theta_C\left( \text{tr}\left(P_{11} J_{q\,L}^\mu P_{32} J_{q\,L\,\mu}  \right)  + \text{tr}\left(P_{11} J_{q\,L}^\mu P_{23} J_{q\,L\,\mu}  \right) \right)\right]\,.
\end{split}
\end{equation}
Where we have defined the projectors $(P_{kl})^i_{\ j}=\delta^i_k\delta_j^l$.

By doing $SU(3)_L\times SU(3)_R$ rotations of equal angle we can derive Ward identities for the vector $SU(3)$ flavor currents. The currents transform as
\begin{equation}
\delta_\theta J_q^\mu=i\theta_A [T^A,J^\mu_q]
\end{equation}

Defining $J_q^\mu=J_{q\,L}^\mu+J_{q\,R}^\mu$ and $J_l^\mu=J_{l\,L}^\mu+J_{l\,R}^\mu$. The covariant derivative under a background flavor gauge potential $A_\mu^f$ is
\begin{equation}
D_\mu J_q^\mu=\partial_\mu J_q^\mu-i[A_\mu^f,J_q^\mu]\,.
\end{equation}
We will take the normalization for the gauge group generators to be $\text{tr}(T^A T^B)=\frac{1}{2}\delta^{AB}$. Then, the Ward identities are
\begin{equation}
\begin{split}
&2\text{tr}\left(D_\mu J_q^{\mu} T^A\right)=\\
&2i\sqrt{2} G_F\left[ \cos^2\theta_C \,\left(\text{tr}\left(P_{11} [T^A, J_{q\,L}^\mu] P_{22} J_{q\,L\,\mu}  \right)+\text{tr}\left(P_{11} J_{q\,L}^\mu P_{22} [T^A,J_{q\,L\,\mu}]  \right)\right)\right.\\
& +\sin^2 \theta_C \,\left( \text{tr}\left(P_{11} [T^A,J_{q\,L}^\mu] P_{33} J_{q\,L\,\mu}  \right)+\text{tr}\left(P_{11}J_{q\,L}^\mu P_{33} [T^A,J_{q\,L\,\mu}]  \right)\right)\\
&+\cos\theta_C \sin\theta_C\left( \text{tr}\left(P_{11} [T^A,J_{q\,L}^\mu] P_{23} J_{q\,L\,\mu}  \right)+\text{tr}\left(P_{11} J_{q\,L}^\mu P_{23} [T^A,J_{q\,L\,\mu}]  \right)\right. \\
&\left.\left. + \text{tr}\left(P_{11} [T^A,J_{q\,L}^\mu] P_{32} J_{q\,L\,\mu}  \right)+ \text{tr}\left(P_{11} J_{q\,L}^\mu P_{32}[T^A, J_{q\,L\,\mu}]  \right) \right)\right]\,.
\end{split}
\end{equation}
Combining the traces and using the cyclic property we have that
\begin{equation}
\begin{split}
&D_\mu J_q^{\mu} =\\
&i\sqrt{2} G_F\left( \cos^2\theta_C \,\left[\,J_{q\,L\,\mu}\,,\,P_{11} J_{q\,L}^\mu P_{22} +P_{22} J_{q\,L}^\mu P_{11}\,  \right]\right.\\
& +\sin^2 \theta_C \,\left[\,J_{q\,L\,\mu}\,,\,P_{11} J_{q\,L}^\mu P_{33} +P_{33} J_{q\,L}^\mu P_{11}\,  \right]\\
&\left.+\cos\theta_C \sin\theta_C\left(  \left[\,J_{q\,L\,\mu}\,,\,P_{11} J_{q\,L}^\mu P_{23} +P_{23} J_{q\,L}^\mu P_{11}\,  \right]+  \left[\,J_{q\,L\,\mu}\,,\,P_{11} J_{q\,L}^\mu P_{32} +P_{32} J_{q\,L}^\mu P_{11}\,  \right]\right)\right)\,.
\end{split}
\end{equation}
In general, there will also be contributions on the right-hand side depending on the quark masses, but they should vanish for diagonal components.

We can do a similar derivation taking into account the leptons if we introduce projectors
\begin{equation}
    (\tilde{P}_{ai})^b_{\ j}=\delta_a^b\delta^i_j,\ \ (\tilde{P}^T_{ia})^j_{\ b}=\delta^a_b\delta^j_i,
\end{equation}
Then
\begin{equation}
    \tilde{P}J^\mu_q\tilde{P}^T\sim J_l^\mu,\ \ \tilde{P}^T J^\mu_l\tilde{P}\sim J_q^\mu. 
\end{equation}
And introduce the $SU(2)_L\times SU(2)_R$ rotations of the leptons, such that the diagonal rotations transform the vector currents as
\begin{equation}
    \delta_\phi J_l^\mu=i\phi_\alpha[\tau^\alpha,J^\mu_l].
\end{equation}
With similar normalization for the generators ${\rm tr}(\tau^\alpha\tau^\beta)=\frac{1}{2}\delta^{\alpha\beta}$.

\section{Linearized equations of motion of fluctuations}\label{app:eoms}

The transverse spatial components of the gauge fields $k^i L_i=k^i R_i=0$ decouple from the other fluctuations. Their equations are
\begin{subequations}
\bea 
0 &=& {L_i}''+ \left(\frac{f'}{f}-\frac{1}{Z}\right){L_i}'+
\left( \frac{\bar\omega^2}{f^2}-\frac{{\bar k}^2}{f}\right)L_i \,,\nonumber\\
\label{eq:lieq}\\
 0 &=& {R_i}''+ \left(\frac{f'}{f}-\frac{1}{Z}\right){R_i}'+
\left( \frac{\bar\omega^2}{f^2}-\frac{{\bar k}^2}{f}\right)R_i \,, \nonumber\\ 
\label{eq:rieq}
\eea 
\end{subequations}
When the quark masses are nonzero, the time $L_0$, $R_0$ and longitudinal $L_\parallel =\frac{k^i}{k} L_i$, $R_\parallel=\frac{k^i}{k} R_i$ components of the gauge fields are coupled to the scalar fluctuations, but they decouple otherwise. The left gauge field equations are
\begin{subequations}
\begin{eqnarray}
 0&=&  {L_\parallel}''+\left(\frac{f'}{f}-\frac{1}{Z}\right){L_\parallel}'+\frac{\bar\omega^2}{f^2}L_\parallel+\frac{\bar k \bar\omega}{f^2}L_0, \label{eq:lkeq}\\
     0&=&  {L_0}''-\frac{1}{Z}{L_0}'-\frac{\bar k^2}{f}L_0-\frac{\bar k \bar\omega}{f}L_\parallel ,\label{eq:l0eq}\\
     0&=&  \bar \omega{L_0}'+\bar k f {L_\parallel}'\,. \label{eq:lconseq}
\end{eqnarray}
\end{subequations}
The right gauge field equations are
\begin{subequations}
\begin{eqnarray}
 0&=&  {R_\parallel}''+\left(\frac{f'}{f}-\frac{1}{Z}\right){R_\parallel}'+\frac{\bar\omega^2}{f^2}R_\parallel+\frac{\bar k \bar\omega}{f^2}R_0,\label{eq:rkeq}\\
     0&=&  {R_0}''-\frac{1}{Z}{R_0}'-\frac{\bar k^2}{f}R_0-\frac{\bar k \bar\omega}{f}R_\parallel,\label{eq:r0eq}\\
     0&=&  \bar \omega{R_0}'+\bar k f {R_\parallel}'\,.\label{eq:rconseq}
\end{eqnarray}
\end{subequations}
The scalar equations are     
\begin{eqnarray}
0 &=& {X}''+\left(\frac{f'}{f}-\frac{3}{Z} \right){X}'+\left(\frac{\bar \omega^2}{f^2}-\frac{\bar k^2}{f}+\frac{3}{Z^2 f}\right)X \;.\label{eq:Xeq}
\end{eqnarray}
We can simplify the equations in the longitudinal sector by introducing the electric fields
\begin{equation}
E_L=\bar k L_0+\bar \omega L_\parallel,\qquad E_R=\bar k R_0+\bar \omega R_\parallel\,.
\end{equation}
The constraints \eqref{eq:lconseq} and \eqref{eq:rconseq} fix the time and longitudinal components of the gauge fields in terms of the electric fields and the scalars
\begin{subequations}\label{eq:timelongsolve}
\begin{eqnarray}
&{L_\parallel}'=\frac{\bar \omega{E_L}'}{\bar \omega^2-\bar k^2 f}\,,\quad &{L_0}'=- \frac{\bar k f{E_L}'}{\bar \omega^2-\bar k^2 f}\,,\\
&{R_\parallel}'=\frac{\bar \omega{E_R}'}{\bar \omega^2-\bar k^2 f}\,,\quad &{R_0}'=- \frac{\bar k f{E_R}'}{\bar \omega^2-\bar k^2 f}\;,
\end{eqnarray}
\end{subequations}
The combinations $\bar\omega f/Z$ \eqref{eq:lkeq} $+$ $\bar k/Z$ \eqref{eq:l0eq}, and  $\bar\omega f/Z$ \eqref{eq:rkeq} $+$ $\bar k/Z$ \eqref{eq:r0eq} produce the equations for the left and right electric fields respectively
\begin{subequations}
\begin{eqnarray}
 0&=&  \partial_Z \left(\frac{f}{Z} \frac{{E_L}'}{\bar \omega^2-\bar k^2 f}\right)+\frac{1}{f Z}E_L, \label{eq:ELeq}\\
0&=&  \partial_Z \left(\frac{f}{Z} \frac{{E_R}'}{\bar \omega^2-\bar k^2 f}\right)+\frac{1}{f Z}E_R\;. \label{eq:EReq}
\end{eqnarray}
\end{subequations}

\section{Solutions in AdS Reissner-Norsdtrom}\label{app:solrn}

We want to solve \eqref{eq:lieq} and \eqref{eq:ELeq} without the terms depending on the masses, in the charged black brane geometry \eqref{eq:blackbranemet} with parameters \eqref{eq:MQAdSRN}. We can solve the equations of motion in the near-extremal $AdS_2$ throat \eqref{eq:AdS2metric} and in the asymptotically $AdS_5$ metric \eqref{eq:blackbranemet}. When $|\bar \omega|\to 0$ the two sets of solutions have an overlapping region $\zeta\to 0$ in the throat and $Z\to 1$ in the outer region. To leading order in the small $|\bar \omega|$ expansion, the equations of motion of the left gauge fields \eqref{eq:lieq} and \eqref{eq:ELeq} in the inner region are
\begin{subequations}\label{eq:innereqs}
\begin{eqnarray}
0 &=& \partial_\zeta^2 L_i +\dfrac{F'}{F}\partial_\zeta L_i +\frac{1}{\zeta^2 F^2}\left(\zeta^2 -\frac{\bar k^2}{12}F\right)L_i\,,\\
 0 &=&  \partial_\zeta^2 E_L + \left( \frac{F'}{F}-\frac{2}{\zeta} \frac{\bar k^2}{12}\frac{1}{\zeta^2-\frac{\bar k^2}{12}F}\right)\partial_\zeta E_L +\frac{1}{\zeta^2 F^2}\left(\zeta^2-\frac{\bar k^2}{12}F\right)E_L .
\end{eqnarray}
\end{subequations}
Outside the $AdS_2$ throat, we can approximate the geometry by the extremal (zero temperature) Reissner-Norstrom solution, with $f(Z)=1-3Z^4+2Z^6$. The equations in the outer region to leading order in the small frequency expansion are
\begin{subequations}\label{eq:outereqs}
\begin{eqnarray}
0 &=& \partial_Z\left(\dfrac{f(Z)}{Z}\partial_Z L_i \right) -\dfrac{\bar k^2}{Z}L_i+O(\bar \omega^2)\,,\\
 0 &=&  \partial_Z\left(\dfrac{\partial_Z E_L}{Z}\right) -\dfrac{\bar{k}^2}{Z\,f(Z)}E_L +O(\bar \omega^2) .
\end{eqnarray}
\end{subequations}

\subsection{Inner solutions of the transverse components}

Let us first find the solutions to the equations in the inner region \eqref{eq:innereqs}. At $T=0$ ($F=1$), there is a simple solution to the equation of the transverse components
\begin{equation}
  L_{i,I}(\zeta) =\sqrt{\zeta}\,\left( C_i^{(1)} \Theta(\omega) H_{\nu_k}^{(1)}\left(\zeta\right)+C_i^{(2)} \Theta(-\omega)  H_{\nu_k}^{(2)}\left(\zeta\right)\right)\;,
\end{equation}
where the coefficients have been chosen so that the solution satisfies ingoing boundary conditions and we have defined
\begin{equation}
\nu_k=\sqrt{\frac{1}{4}+\frac{\bar{k}^2}{12}}\;.
\end{equation}
These solutions should match with the solutions of the outer region for $\zeta\to 0$. The leading terms in the expansion are
\begin{equation}\label{eq:LIexp}
  L_{i,I}(\zeta)\; \sim \; A_i \zeta^{\frac{1}{2}-\nu_k}+B_i \zeta^{\frac{1}{2}+\nu_k}\;,
\end{equation}
where
\begin{subequations}
\begin{eqnarray}
A_i &=&   -i\frac{2^{\nu_k} \Gamma(\nu_k)}{\pi} \left(C_i^{(1)}\Theta(\omega)-C_i^{(2)}\Theta(-\omega)\right) ,\\
B_i &=& \frac{1}{2^{\nu_k} \Gamma(\nu_k+1)}\left(C_i^{(1)}\Theta(\omega)+C_i^{(2)}\Theta(-\omega)+i\cot(\pi \nu_k)(C_i^{(1)}\Theta(\omega)-C_i^{(2)}\Theta(-\omega))  \right)\;.
\end{eqnarray}
\end{subequations}
In terms of the original coordinate $Z$, related to $\zeta$ through \eqref{eq:zetacoord}, the expansion \eqref{eq:LIexp} is
\begin{equation}\label{eq:Lizeta0}
  L_{i,I}(\zeta)\; \sim \; A_i\left(\frac{|\bar\omega|}{12} \right)^{\frac{1}{2}-\nu_k} \left[(1-Z)^{-\frac{1}{2}+\nu_k}+\mathcal{G}_L(\bar\omega,\bar k) (1-Z)^{-\frac{1}{2}-\nu_k}\right)\;,
\end{equation}
where the IR Green's function is 
\begin{equation}
\mathcal{G}_L(\bar\omega,\bar k)=\dfrac{\pi }{2^{2\nu_k}\Gamma\left(\nu_k\right)\Gamma\left(1+\nu_k\right)}\left(-\cot(\pi \nu_k)+i\, \text{sign}(\omega)\right)\left(\frac{|\bar \omega|}{12}\right)^{2\nu_k}.
\end{equation}
At non-zero temperatures, it is still possible to find analytic solutions
\begin{equation}
  L_{i,I}(\zeta) = A_i \varphi_{\nu_k}^{-}\left(\zeta\right)+B_i\varphi_{\nu_k}^{+}\left(\zeta\right)\;,
\end{equation}
where
\begin{equation}\label{eq:Linsol}
\varphi_{\nu_k}^\pm(\zeta)=\left(1-\frac{\zeta^2}{\zeta_0^2}\right)^{-\frac{i}{2}\text{sign}(\omega)\zeta_0} \left(\frac{\zeta}{\zeta_0}\right)^{\frac{1}{2}\pm \nu_k}{}_2 F_1\left(\frac{1}{4}-\frac{i}{2}\text{sign}(\omega)\zeta_0\pm \frac{\nu_k}{2}\,,\, \frac{3}{4}-\frac{i}{2}\text{sign}(\omega)\zeta_0\pm \frac{\nu_k}{2}\,,\, 1\pm \nu_k\,;\,\frac{\zeta^2}{\zeta_0^2} \right)\;.
\end{equation}
The condition that the solution is ingoing fixes
\begin{equation}
B_i=\frac{\Gamma \left(-\nu_k\right) \Gamma \left(\frac{1}{2}-i \,\text{sign}(\omega)\zeta_0+\nu_k\right)}{2^{2\nu_k}\Gamma \left(\nu_k\right) \Gamma \left(\frac{1}{2}-i \,\text{sign}(\omega) \zeta_0-\nu_k\right)}  A_i\;.
\end{equation}
Using that $Z_*\approx 1$, we arrive at 
\begin{equation}\label{eq:LizetaT}
  L_{i,I}(\zeta)\; \sim \; A_i\left(\frac{|\bar\omega|}{12 \zeta_0} \right)^{\frac{1}{2}-\nu_k} \left[(1-Z)^{-\frac{1}{2}+\nu_k}+\mathcal{G}_L(\bar\omega,\bar k,\zeta_0) (1-Z)^{-\frac{1}{2}-\nu_k}\right)\;,
\end{equation}
where now
\begin{equation}
\mathcal{G}_L(\bar\omega,\bar k,\zeta_0)=\frac{\Gamma \left(-\nu_k\right) \Gamma \left(\frac{1}{2}-i \,\text{sign}(\omega)\zeta_0+\nu_k\right)}{2^{2\nu_k}\Gamma \left(\nu_k\right) \Gamma \left(\frac{1}{2}-i \,\text{sign}(\omega) \zeta_0-\nu_k\right)}\left(\frac{|\bar\omega|}{12 \zeta_0} \right)^{2\nu_k}.
\end{equation}
In terms of temperature the IR Green's function is approximately
\begin{equation}
\mathcal{G}_L(\bar\omega,\bar k,\bar T)\approx \frac{\Gamma \left(-\nu_k\right) \Gamma \left(\frac{1}{2}-i \,\frac{\omega}{2\pi T}+\nu_k\right)}{2^{2\nu_k}\Gamma \left(\nu_k\right) \Gamma \left(\frac{1}{2}-i \,\frac{ \omega}{2\pi T}-\nu_k\right)}\left(\frac{\pi \bar T}{6} \right)^{2\nu_k}.
\end{equation}
For frequencies smaller than the temperature, we can expand the IR Green's function as
\begin{equation}\label{eq:GLsmallw}
\mathcal{G}_L(\bar\omega,\bar k,\bar T)\approx \frac{\Gamma \left(-\nu_k\right) \Gamma \left(\frac{1}{2}+\nu_k\right)}{2^{2\nu_k}\Gamma \left(\nu_k\right) \Gamma \left(\frac{1}{2}-\nu_k\right)}\left(\frac{\pi \bar T}{6} \right)^{2\nu_k}\left(1-\frac{i}{2}\tan(\pi \nu_k)\frac{\omega}{T}\right)+O\left(\frac{\omega^2}{T^2}\right).
\end{equation}
The imaginary term produces the leading contribution to the spectral function. We can obtain these first terms in the $\omega/T$ by first extracting the ingoing factor from the solutions
\begin{equation}
L_{i,I}(\zeta)=\left(1-\frac{\zeta^2}{\zeta_0^2}\right)^{-\frac{i}{2}\text{sign}(\omega)\zeta_0} \psi_i(\zeta),
\end{equation}
and then expanding the equations in powers of $\zeta_0$ and solving order by order. Changing variables to $u=\zeta^2/\zeta_0^2$, the equations for the first two orders $\psi_i=\psi_i^0+\zeta_0\psi_i^1+\cdots$ are
\begin{subequations}
\begin{eqnarray}
 0&=& \partial_u\left(\sqrt{u}(1-u)\partial_u \psi_i^0\right)+\frac{1-4\nu_k^2}{16 u^{3/2}}\psi_i^0\,,\\
 0&=& \partial_u\left(\sqrt{u}(1-u)\partial_u \psi_i^1\right)+\frac{1-4\nu_k^2}{16 u^{3/2}}\psi_i^1+i\,\text{sign}(\omega)\sqrt{u} \left(\partial_u\psi_i^0+\frac{1}{4 u} \psi_i^0 \right)\;.
\end{eqnarray}
\end{subequations}
The solutions to zeroth order are
\begin{equation}
\psi_i^0(u)= A_i^0 \varphi_{\nu_k}^{0,-}\left(u\right)+B_i\varphi_{\nu_k}^{0,+}\left(u\right),
\end{equation}
with
\begin{equation}
\varphi_{\nu_k}^{0,\pm}(u)= u^{\frac{1}{4}(1\pm 2\nu_k)}{}_2 F_1\left(\frac{1}{4}\pm \frac{\nu_k}{2}\,,\, \frac{3}{4}\pm \frac{\nu_k}{2}\,,\, 1\pm \nu_k\,;\,u \right)\;.
\end{equation}
Imposing regularity of this solution when $u\to 1$ ($\zeta\to \zeta_0$), i.e. removing logarithmic terms, fixes the relation between the coefficients
\begin{equation}
B_i^0=\frac{\Gamma \left(-\nu_k\right) \Gamma \left(\frac{1}{2}+\nu_k\right)}{2^{2\nu_k}\Gamma \left(\nu_k\right) \Gamma \left(\frac{1}{2}-\nu_k\right)}  A_i^0\;.
\end{equation}
The next-order solution can be obtained using a Green's function. First we identify the subleading solution when $u\to 0$ ($\zeta\to 0$), $\varphi^<$, and the ingoing solution when $u\to 1$, $\varphi^>$,
\begin{equation}
\varphi^<(u)=\varphi_{\nu_k}^{0,+}(u)\,,\quad \varphi^>(\zeta)=\varphi_{\nu_k}^{0,-}(u)+\frac{\Gamma \left(-\nu_k\right) \Gamma \left(\frac{1}{2}+\nu_k\right)}{2^{2\nu_k}\Gamma \left(\nu_k\right) \Gamma \left(\frac{1}{2}-\nu_k\right)} \varphi_{\nu_k}^{0,+}(u)\;.
\end{equation}
With these and the Wronskian of the two subleading and ingoing functions we construct a Green's function that maintains the boundary conditions of the leading order solution 
\begin{equation}
G(u,u')=-\frac{1}{\nu_k}\left\{\begin{array}{lcl}  \varphi^<(u)\varphi^>(u') &,& u<u' \\   \varphi^>(u)\varphi^<(u') &,& u>u' \end{array} \right.
\end{equation}
Then, taking into account that $\psi_i^0=A_i^0\varphi^>(u)$, the subleading solution is
\begin{equation}
\psi_i^1(u)=-i\,\text{sign}(\omega)A_i^0\int_0^1 du' G(u,u')\sqrt{u'} \left(\partial_{u'}\varphi^>(u')+\frac{1}{4 u'} \varphi^>(u') \right)\;.
\end{equation}
We can now expand the leading and subleading solutions for $u\to 0$. One can check that all contributions from $\psi_i^1$ are subleading compared to the leading term $\sim u^{\frac{1}{4}(1-2 \nu_k)}$. Also, there is just one contribution to the term $\sim u^{\frac{1}{4}(1+2 \nu_k)}$, from a total derivative 
\begin{equation}
\int_0^1 du' G(u,u')\sqrt{u'} \left(\partial_{u'}\varphi^>(u')+\frac{1}{4 u'} \varphi^>(u') \right)\sim \varphi^<(u)\lim_{u\to 0}\int_u^1 du' \partial_{u'}\left(\sqrt{u'} \frac{(\varphi^>(u'))^2}{2}\right)=\varphi^<(u)  \frac{(\varphi^>(1))^2}{2}\;.
\end{equation}
Taking this into account, the relevant terms in the expansion are
\begin{eqnarray}
\psi_i^0 &\sim & \left( u^{\frac{1}{4}(1- 2\nu_k)}+\frac{\Gamma \left(-\nu_k\right) \Gamma \left(\frac{1}{2}+\nu_k\right)}{2^{2\nu_k}\Gamma \left(\nu_k\right) \Gamma \left(\frac{1}{2}-\nu_k\right)}  u^{\frac{1}{4}(1+2 \nu_k)}\right)A_i^0 \,,\\
\psi_i^1 &\sim & -i\,\text{sign}(\omega)\pi \tan(\pi \nu_k) \frac{\Gamma \left(-\nu_k\right) \Gamma \left(\frac{1}{2}+\nu_k\right)}{2^{2\nu_k}\Gamma \left(\nu_k\right) \Gamma \left(\frac{1}{2}-\nu_k\right)}  u^{\frac{1}{4}(1+2 \nu_k)}  A_i^0\;.
\end{eqnarray}
It is straightforward to check that this reproduces the first two terms in the $\omega/T$ expansion of the  IR Greens' function \eqref{eq:GLsmallw}.

\subsection{Inner solutions of the longitudinal components}

The equation for the electric field in \eqref{eq:innereqs} does not have a simple analytic solution, but since we are interested in frequencies $\omega/T\lesssim 1$, we can apply the same expansion as for the transverse components to find the leading contribution to the spectral function.

We first extract the ingoing factor from the solutions
\begin{equation}
E_{L,I}(\zeta)=\left(1-\frac{\zeta^2}{\zeta_0^2}\right)^{-\frac{i}{2}\text{sign}(\omega)\zeta_0} \psi_i(\zeta),
\end{equation}
and then expand the equations in powers of $\zeta_0$ and solve them order by order. Changing variables to $u=\zeta^2/\zeta_0^2$, the equations for the first two orders $\psi_i=\psi_i^0+\zeta_0\psi_i^1+\cdots$ are
\begin{subequations}
\begin{eqnarray}
 0&=& \partial_u\left(u^{3/2}\partial_u \psi_i^0\right)+\frac{1-4\nu_k^2}{16 u^{1/2}(1-u)}\psi_i^0\,,\\
 0&=& \partial_u\left(u^{3/2}\partial_u \psi_i^1\right)+\frac{1-4\nu_k^2}{16 u^{1/2}(1-u)}\psi_i^1+i\,\text{sign}(\omega)\frac{u^{3/2}}{1-u} \left(\partial_u\psi_i^0+\frac{3-u}{4 u(1-u)} \psi_i^0 \right)\;.
\end{eqnarray}
\end{subequations}
The solutions to zeroth order are
\begin{equation}
\psi_i^0(u)= A_i^0 \varphi_{\nu_k}^{0,-}\left(u\right)+B_i\varphi_{\nu_k}^{0,+}\left(u\right),
\end{equation}
with
\begin{equation}\label{eq:Einsol}
\varphi_{\nu_k}^{0,\pm}(u)= u^{-\frac{1}{4}(1\mp 2 \nu_k)}{}_2 F_1\left(-\frac{1}{4}\pm \frac{\nu_k}{2}\,,\, \frac{1}{4}\pm \frac{\nu_k}{2}\,,\, 1\pm \nu_k\,;\,u \right)\;.
\end{equation}
Imposing regularity of this solution when $u\to 1$ ($\zeta\to \zeta_0$), i.e. removing logarithmic terms, fixes the relation between the coefficients
\begin{equation}
B_i^0=\frac{\Gamma \left(-\nu_k\right) \Gamma \left(\frac{3}{2}+\nu_k\right)}{2^{2\nu_k}\Gamma \left(\nu_k\right) \Gamma \left(\frac{3}{2}-\nu_k\right)}  A_i^0\;.
\end{equation}
It turns out that this makes the solution vanish at the horizon, something that will be relevant below.

The next-order solution can be obtained in terms of a Green's function. First, we identify the subleading solution when $u\to 0$ ($\zeta\to 0$), $\varphi^<$, and the ingoing solution when $u\to 1$, $\varphi^>$,
\begin{equation}
\varphi^<(u)=\varphi_{\nu_k}^{0,+}(u)\,,\quad \varphi^>(\zeta)=\varphi_{\nu_k}^{0,-}(u)+\frac{\Gamma \left(-\nu_k\right) \Gamma \left(\frac{3}{2}+\nu_k\right)}{2^{2\nu_k}\Gamma \left(\nu_k\right) \Gamma \left(\frac{3}{2}-\nu_k\right)} \varphi_{\nu_k}^{0,+}(u)\;.
\end{equation}
With these and the Wronskian of the two subleading and ingoing functions we construct a Green's function that maintains the boundary conditions of the leading order solution 
\begin{equation}
G(u,u')=-\frac{1}{\nu_k}\left\{\begin{array}{lcl}  \varphi^<(u)\varphi^>(u') &,& u<u' \\   \varphi^>(u)\varphi^<(u') &,& u>u' \end{array} \right.
\end{equation}
Then, taking into account that $\psi_i^0=A_i^0\varphi^>(u)$, the subleading solution is
\begin{equation}
\psi_i^1(u)=-i\,\text{sign}(\omega)A_i^0\int_0^1 du' G(u,u')\frac{{u'}^{3/2}}{1-u'} \left(\partial_{u'}\varphi^>(u')+\frac{3-u'}{4 u'(1-u')} \varphi^>(u') \right)\;.
\end{equation}
We can now expand the leading and subleading solutions for $u\to 0$. One can check that all contributions from $\psi_i^1$ are subleading compared to the leading term $\sim u^{-\frac{1}{4}(1+2 \nu_k)}$. Also, there is just one potential contribution to the term $\sim u^{-\frac{1}{4}(1-2 \nu_k)}$, from a total derivative 
\begin{equation}
\begin{split}
&\int_0^1 du' G(u,u')\frac{{u'}^{3/2}}{1-u'} \left(\partial_{u'}\varphi^>(u')+\frac{3-u'}{4 u'(1-u')} \varphi^>(u') \right)\\
&\qquad \sim \varphi^<(u)\lim_{u\to 0}\int_u^1 du' \partial_{u'}\left(\frac{{u'}^{3/2}}{1-u'} \frac{(\varphi^>(u'))^2}{2}\right)=\varphi^<(u)  \lim_{u'\to 1} \frac{(\varphi^>(u'))^2}{2(1-u')}=0\;.
\end{split}
\end{equation}
Taking this into account, the relevant terms in the expansion are
\begin{eqnarray}
\psi_i^0 &\sim & \left( u^{-\frac{1}{4}(1+2 \nu_k)}+\frac{\Gamma \left(-\nu_k\right) \Gamma \left(\frac{3}{2}+\nu_k\right)}{2^{2\nu_k}\Gamma \left(\nu_k\right) \Gamma \left(\frac{3}{2}-\nu_k\right)}  u^{-\frac{1}{4}(1-2 \nu_k)}\right)A_i^0 \,,\\
\psi_i^1 &\sim & 0 \times  A_i^0  u^{-\frac{1}{4}(1-2 \nu_k)}\;.
\end{eqnarray}
Thus, the leading $\omega/T$ correction actually vanishes for the longitudinal components. This implies that for the calculation of the spectral function, it is enough to compute the transverse contribution.

\subsection{Outer solutions of the longitudinal components}

Although we will not need these solutions to compute the leading contributions to the rates, it will be useful to study them as they can be found analytically.

We are looking for solutions of the electric field in \eqref{eq:outereqs}. First we will change coordinates $u=Z^2$, so the equations takes the form of a Schroedinger equation
\begin{equation}\label{eq:eqELO}
        E_{L,O}''(u)-\dfrac{\bar k^2}{4u(2u+1)(u-1)^2}E_{L,O}=0\;.
\end{equation}
We can transform the above equation into a hypergeometric equation by the following change of coordinate and field redefinition
\begin{align}
        &E_{L,O}(u) = (1+2u)^{c_\pm}(1-u)^{b_\pm}t_\pm(u),\,\,\,\,\,\,c_\pm = \dfrac{1}{2}\pm\nu_k,\,\,\,\,\,b_\pm = 1-c_\pm = \dfrac{1}{2}\mp\nu_k,\\
        &x=\dfrac{3u}{1+2u},\,\,\,\,\,\,\,\,\,\,\,u =\dfrac{x}{3-2x}.
\end{align}
 Then, we get the following hypergeometric equation
    \begin{equation}
        x(1-x)t_\pm''(x)+(\pm 2\nu_k -1)x\,t_\pm'(x) -\dfrac{k^2}{12}\,t_\pm(x)=0.
    \end{equation}
    We can compare the above equation to the standard form $x(1-x)t_\pm''(x) + (c_\pm-(a_\pm+b_\pm+1)x)t_\pm'(x) -a_\pm b_\pm t(x) =0$, to get
\begin{equation}
    c_\pm = 0,\,\,\,\,\,\,\,\, a_\pm = \Bigg\lbrace \frac{1}{2}\mp \nu_k,-\frac{1}{2}\mp \nu_k\Bigg\rbrace,\,\,\,\,\,b_\pm = \Bigg\lbrace-\frac{1}{2}\mp \nu_k,\frac{1}{2}\mp \nu_k\Bigg\rbrace.
\end{equation}
A general solution can be written as a linear combination of a normalizable and a non-normalizable mode at the boundary. In terms of the original radial variable, they have the form
\begin{align}
    &E_{O}^{(0)} = \alpha\, \phi_{NN}(Z) + \beta \,\phi_N(Z),\\ \nonumber
    &\phi_{NN}(Z) = (1-Z^2)^{\frac{1}{2}+\nu_k}(1+2Z^2)^{\frac{1}{2}-\nu_k}\bigg[\bigg(\left(\dfrac{1}{2}-2\nu_k^2\right)H_{-\frac{1}{2}+\nu_k}-\nu_k^2(-1+\log 3)+\\ \nonumber
    &\,\,\,\,\,\,\,\,\,\,+ \dfrac{1}{12}(1+3\log 3)\bigg)\dfrac{3Z^2}{1 + 2Z^2}\,_2F_1\left(\dfrac{1}{2}+\nu_k,\dfrac{3}{2}+\nu_k,2,\dfrac{3Z^2}{1 + 2Z^2}\right)+\\
    &\,\,\,\,\,\,\,\,\,\,+\dfrac{1}{\Gamma\left(-\frac{1}{2}+\nu_k\right)\Gamma\left(\frac{1}{2}+\nu_k\right)}G_{2,2}^{2,2}\left(
\begin{array}{c}
\dfrac{3Z^2}{1+2Z^2}
\end{array}\middle\vert
\begin{array}{c}
\frac{1}{2}-\nu_k,\dfrac{3}{2}-\nu_k\\
0,1\\
\end{array}
\right)\bigg],\\
&\phi_N(Z) = (1-Z^2)^{\frac{1}{2}+\nu_k}(1+2Z^2)^{\frac{1}{2}-\nu_k}\dfrac{3Z^2}{1+2Z^2}\,_2F_1\left(\dfrac{1}{2}+\nu_k,\dfrac{3}{2}+\nu_k,2,\dfrac{3Z^2}{1 + 2Z^2}\right),
\end{align}
where $H_{-\frac{1}{2}+\nu_k}$ is the $(-\frac{1}{2}+\nu_k)^{\text{th}}$ harmonic number and $G$ is the Meijer function.

The non-normalizable and normalizable solutions have been chosen so that their near-boundary expansion is
\begin{align}
\phi_{NN}(Z) = 1 + \dfrac{3}{2}(-1+4\nu_k^2)Z^2\log Z + O(Z^4\log Z),\,\,\,\,\,\,\,\,\,\,\,\,\,\,\,\phi_N(Z) = Z^2 + O(Z^4)\,.
\end{align}
We want to determine the coefficients $\alpha$ and $\beta$ for solutions with a selected behaviour close to the horizon $Z\to 1$. The expansion close to the horizon is 
\begin{equation}
         E_{L,O}(Z) \sim A\left(1-Z\right)^{\frac{1}{2}+\nu_k} + B\left(1-Z\right)^{\frac{1}{2}-\nu_k} +...\,\,\,\,\,\,\,\, Z\to 1,
\end{equation}
These match with the expansions of the $\varphi_{\nu_k}^{0,\pm}$ inner solutions \eqref{eq:Einsol} in the $\zeta\to 0$ limit.
By requiring that $A = 1$ and $B = 0$ we then determine the value of $\alpha$ and $\beta$ for the minus mode while, if we impose $A = 0$ and $B = 1$ we find the coefficients for the plus mode. They turn out to be
\begin{align}
    &\alpha_{\mp} =\dfrac{6^{\pm\nu_k -\frac{1}{2}} \Gamma (\pm\nu_k +1)}{\sqrt{\pi } \,\Gamma \left(\pm\nu_k +\frac{3}{2}\right)},\\
    &\beta_{-} = -\dfrac{2^{\nu_k -\frac{5}{2}} 3^{\nu_k -\frac{1}{2}} \nu_k  \Gamma (\nu_k ) \left(\left(6-24 \nu_k ^2\right) H_{\nu -\frac{1}{2}}-12 \nu_k ^2 (1+\log (3))+1+\log (27)\right)}{\sqrt{\pi }\, \Gamma \left(\nu_k +\frac{3}{2}\right)},\\
    &\beta_+ = -\dfrac{ \pi \nu_k  \left(\csc (2 \pi  \nu_k ) \left(\left(6-24 \nu_k ^2\right) H_{\nu_k -\frac{1}{2}}-12 \nu_k ^2 (\log (3)-1)+1+\log (27)\right)+3 \pi  \left(4 \nu_k ^2-1\right) \sec ^2(\pi  \nu_k )\right)}{2^{-\nu_k +\frac{3}{2}} 3^{\nu_k +\frac{1}{2}}\Gamma \left(\frac{1}{2}-\nu_k \right) \Gamma \left(\frac{3}{2}-\nu_k \right) \Gamma (2 \nu_k +1)}\,.
\end{align}

Since in order to compute the rates we have to perform an integral over all momentum, it is interesting to ascertain the behaviour of the spectral function at large spatial momentum, which will be in part determined by the behaviour of the connection coefficients. From the analytic formulas above one can see that, for $\nu_k\to \infty$, their asymptotic form is
\begin{align}\label{eq:asympalphabeta}
    &\alpha_- \sim 6^{\nu_k-\frac{1}{2}}(\pi \nu_k)^{-1/2},\quad \alpha_+ \sim -6^{-\nu_k-\frac{1}{2}}(\pi \nu_k)^{-1/2}\cot(\pi\nu_k) ,\\
    &\beta_{-} \sim 6^{\nu_k+\frac{1}{2}}\nu_k^{3/2}\log\left(3^{1/2}e^{-1/2}e^{\gamma_E}\nu_k\right),\\
    &\beta_+ \sim  6^{-\nu_k+\frac{1}{2}}\nu_k^{3/2} \sqrt{\pi}\left(1-\frac{\cot(\pi \nu_k)}{\pi} \log\left(3^{1/2}e^{-1/2}e^{\gamma_E}\nu_k\right)\right)\,.
\end{align}
We can recover this behaviour of the connection coefficients by using a WKB approximation at large spatial momentum. The equation of motion \eqref{eq:eqELO} is already in Schroedinger form with a potential
\begin{equation}\label{eq:UEpot}
U_E(u)=\dfrac{\bar k^2}{4u(2u+1)(u-1)^2}\;.
\end{equation}
We expand the solutions as
\begin{align}
    &E_{L,O}(u) = C\exp\left(\bar{k} \,S_{0}(u) + S_{1}(u) + O(1/\bar{k})\right).
\end{align}
The equation for the leading term in the $1/\bar{k}$ expansion is
\begin{equation}
    S_0'(u)^2-\dfrac{1}{8 u^4-12 u^3+4 u} = 0,
\end{equation}
which has an analytical solution
\begin{equation}
    S_0(u) = \pm\dfrac{1}{\sqrt{3}}\tanh ^{-1}\left( \sqrt{\dfrac{3u}{2u+1}}\right) + c_0.
\end{equation}
Then, the subleading term is given by
\begin{equation}
    S_1(u) = \log\left(\sqrt{2}\,\sqrt{1- u} \,\sqrt[4]{u (1+2u)}\right) + c_1.
\end{equation}
Then the solution up to this order reads
\begin{equation}
    E_{L,O}(u)\approx \psi(u)= C_\pm \,\sqrt{1- u} \,\sqrt[4]{u (1+2u)}\exp\left(\pm\dfrac{\bar{k}}{\sqrt{3}}\tanh ^{-1}\left( \sqrt{\dfrac{3u}{1+2u}}\right)\right).
\end{equation}
The WKB solution near the boundary has the following expansion
\begin{equation*}
    \psi(u)\sim C_\pm u^{1/4}e^{\pm k\sqrt{u}}\,\,\,\,\,\,\,\,\,\,u\to 0.
\end{equation*}
Note that the two terms in the solution directly correspond to the "plus" and "minus" modes at the horizon
\begin{equation*}
    \psi(u) \sim \widetilde{C}_\pm \left(\dfrac{1}{1-u}\right)^{\pm\bar{k}/\sqrt{12}}\,\,\,\,\,\,\,\,\,u\to 1.
\end{equation*}
We have to fix the overall coefficient to properly normalize the WKB solutions in the original coordinate $z$ which are
\begin{align}
   &E(z) = C_\pm \,\sqrt{1- z^2} \,\sqrt{z}\,(1+2z^2)^{1/4}\exp\left(\pm\dfrac{\bar{k}}{\sqrt{3}}\tanh ^{-1}\left( \sqrt{\dfrac{3z^2}{1+2z^2}}\right)\right).
\end{align}
Then, the overall coefficients are found to be
\begin{align}
    &C_\pm = 6^{\mp \frac{\bar{k}}{\sqrt{12}}}.
\end{align}
Now we have to match these results with the solutions close to the boundary and read the boundary coefficients. Close to the boundary, the potential is to leading order
\begin{equation*}
  U_E(u) \sim \dfrac{\bar{k}^2}{4u}.
\end{equation*} 
A solution for this approximated potential can be given in terms of modified Bessel functions
\begin{equation}
    \psi(u)\approx c_1 \sqrt{u} \,K_1\left(\bar{k} \,\sqrt{u}\right)+ c_2\sqrt{u} \,I_1\left(\bar{k}\,\sqrt{u}\right).
\end{equation}
The expansion for large $\bar{k}$ is
\begin{equation*}
    \sqrt{u}\,K_1\left(\bar{k} \,\sqrt{u}\right)\sim \sqrt{\dfrac{\pi }{2\bar{k}}}u^{1/4}e^{-\bar{k}\sqrt{u}},\,\,\,\,\,\,\,\,\,\,\,\,\,\,\,\, \sqrt{u}\,I_1\left(\bar{k} \,\sqrt{u}\right)\sim\dfrac{u^{1/4}}{\sqrt{2\pi\bar{k}}}e^{\bar{k}\sqrt{u}},
\end{equation*}
which is consistent with the WKB solution. We fix the coefficient $c_{1,2}$ with the formulas for $C_\pm$ so that the WKB solution for $z\to 0$ and the solution close to the boundary for large $\bar{k}$ coincide. Finally, we can read off from the $z\to 0$ expansion the coefficients $\alpha_\pm$ and $\beta_\pm$ which are respectively the coefficients of the non-normalizable and normalizable mode. In a compact form, they are
\begin{align}\label{eq:WKBalphabeta}
    &\alpha_\pm \sim \mp 6^{\mp\frac{\bar{k}}{\sqrt{12}}}\bar{k}^{-1/2} ,\\
    &\beta_\pm \sim \mp 6^{\mp\frac{\bar{k}}{\sqrt{12}}}\,\bar{k}^{3/2}\,\log\bar{k}.
\end{align}
Comparing to \eqref{eq:asympalphabeta}, and taking into account $\nu_k\sim \frac{\bar k}{\sqrt{12}}$ in this limit, the WKB approximation captures the leading $\bar k$ dependence up to the oscillatory factors.

\subsection{Outer solutions of the transverse components}

We are looking for solutions of the transverse components in \eqref{eq:outereqs}. We do not have a close analytic solution in this case, so we must resort to numerics. The general solution is a linear combination of two modes $\eta_{L,\mp}(Z)$ which when $Z\to 1$ match with the inner solutions \eqref{eq:Linsol} for $\zeta\to 0$,
\begin{equation}
    \eta_{L,\pm}(Z) \sim \left(1-Z\right)^{-\frac{1}{2}\mp\nu_k} + ... \,\,\,\,\,\,\,\, Z\to 1.
\end{equation}
There is another basis of solutions we can use to write a general solution, consisting of the normalizable and non-normalizable modes, with a boundary expansion at $Z\to 0$
\begin{align}
\phi_{NN}(Z) = 1 + \dfrac{3}{2}(-1+4\nu_k^2)Z^2\log Z + O(Z^4\log Z),\,\,\,\,\,\,\,\,\,\,\,\,\,\,\,\phi_N(Z) = Z^2 + O(Z^4)
\end{align}
The change of basis is determined by the connection coefficients $\alpha_\pm,\beta_\pm$:
\begin{equation}
    \eta_{L,\pm}(Z) = \alpha_\pm\phi_{NN}(Z) +\beta_\pm\phi_N(Z).
\end{equation}
The connection coefficients can be determined from the solutions by matching the $\eta_{L\,\pm}$ solutions and their first derivative with a linear combination of the normalizable and non-normalizable solutions at any intermediate point $0<\hat{Z}<1$:
\begin{equation}
\begin{pmatrix}
\alpha_\pm \\
\beta_\pm
\end{pmatrix}
=
\dfrac{1}{W(Z)}\begin{pmatrix}
-\partial_Z\phi_{N}(Z) & \phi_{N}(Z) \\
\partial_Z\phi_{NN}(Z) & -\phi_{NN}(Z)
\end{pmatrix}
\begin{pmatrix}
\eta_{L,\pm}(Z) \\
\partial_Z\eta_{L,\pm}(Z)
\end{pmatrix}\bigg|_{Z = \hat{Z}}\,,
\end{equation}
where $W(Z) = \phi_{N}(Z)\partial_Z\phi_{NN}(Z)-\partial_Z\phi_{N}(Z)\phi_{NN}(Z)$.  We will obtain the connection coefficients using this formula, by finding numerically the $\eta_{L\,\pm}$ solutions using shooting from the horizon, and the $\phi_{N},\phi_{NN}$ solutions using shooting from the boundary. We use $\hat{Z}=0.2$ as the matching point, but we have checked that changing this value through the interval of numerical integration does not affect the result within our numerical precision.

For the shooting from the boundary, we introduce a cutoff at $Z=10^{-5}$ and use a series expansion to give initial values for the solution and its first derivative at the cutoff. The expansion is 
\begin{align}
    &\phi_{N}(Z) = \sum\limits_{n = 1}^{N_a} a_n Z^{2n},\\
    &\phi_{NN}(Z) = b_0 + \sum\limits_{n = 2}^{N_b} b_n Z^{2n} +\log(Z)\sum\limits_{n = 1}^{N_c} c_n Z^{2n}.
\end{align}
Where we take $N_a=8$, $N_b=6$ and $N_c=7$. The first coefficients of the expansion given that $a_1 = b_0 = 1$ are
\begin{align}
    &a_2 = \dfrac{\bar{k}^2}{8},\,\,\,\,a_3 = 1 + \dfrac{\bar{k}^4}{192},\,\,\,\,a_4 = -\dfrac{1}{2} + \dfrac{5\bar{k}^2}{24} + \dfrac{\bar{k}^6}{9216},\,...\\
    &b_2 = -\dfrac{3\bar{k}^4}{64},\,\,\,\,b_3 = \dfrac{\bar{k}^2}{6} - \dfrac{7\bar{k}^6}{2304}\,\,...\,\,\,\,\,\,c_1 =\dfrac{\bar{k}^2}{2},\,\,\,\,c_2 = \dfrac{\bar{k}^4}{16},\,\dots
\end{align}

For the shooting from the horizon we introduce a cutoff at $Z=1-10^{-5}$. We extract the leading power in the horizon expansion from the solutions
\begin{equation}
    \eta_{L,\mp}(Z) = \left(1-Z\right)^{-\frac{1}{2}\pm\nu_k}p_{\mp}(Z)\;.
\end{equation}
The equation of motion we solve is
\begin{align}
    &\partial_Z^2 p_\mp(Z) +\left(\dfrac{1+2Z\left(-1 \pm\nu_k\right)}{Z(-1+Z)} + \dfrac{f'(Z)}{f(Z)}\right)\partial_Z p_\mp(Z)+\\ \nonumber
    &+\dfrac{(\pm \nu_k-1)\left[(2-5Z \pm 2Z\nu_k)f(Z) + 2(1-Z)Z\left(6(-1+Z)(1\pm 2\nu_k) -f'(Z)\right)\right]}{4 (Z-1)^2 Z f(Z)}\,p_\mp(Z).
\end{align}
To give initial values at the cutoff we use the series expansion 
\begin{align}
    p_\mp = \sum\limits_{n=0}^{N_j} j_{\mp,n}\left(1-Z\right)^n,
\end{align}
and shooting it from the horizon towards the boundary. We expand up to $N_j=8$ terms, with the first coefficients in the expansion, given that $j_{\mp,0}=1$, are
\begin{align}\label{eq:coefj}
    j_{\mp,1} = \dfrac{8\sqrt{3}\sqrt{3+\bar{k}^2} \mp 24 \pm 7\bar{k}^2}{12(\sqrt{3}\sqrt{3+\bar{k}^2} \pm 3)},\,\,\,\,j_{\mp,2} = \dfrac{\bar{k}^2 \left(49 \sqrt{3} \sqrt{\bar{k}^2+3}\pm534\right)-156 \left(\sqrt{3} \sqrt{\bar{k}^2+3}\mp3\right)}{864 \left(\sqrt{3} \sqrt{\bar{k}^2+3}\pm6\right)}, \dots
\end{align}

Although we do not have an analytic formula for the connection coefficients, we can obtain analytic expressions for their asymptotic behaviour at large spatial momentum, by applying the same WKB approximation we used for the coefficients of the longitudinal components. First, we should put the equation of motion for $L_i$ in \eqref{eq:outereqs} in Schroedinger form. This is achieved by the change of variables $u=Z^2$ followed by the redefinition
\begin{equation}
L_i(u)= \left(1-u\right) \left(1+2 u\right)^{-1/2}\psi_{L_i}(u)\;.
\end{equation}
The equation of motion is now
\begin{equation}
\psi_{L_i}''(u)-U_{L_i}(u)\psi_{L_i}(u) = 0\,, \qquad U_{L_i}(u) = \frac{k^2}{4 (1-u)^2 u (2 u+1)}-\frac{3 (u+1)}{(1-u) (2 u+1)^2}\;.
\end{equation}
At large $\bar k$ the potential is to leading order the same as for the longitudinal modes \eqref{eq:UEpot}
\begin{equation}
U_{L_i}(u)\sim \frac{k^2}{4 (1-u)^2 u (2 u+1)}=U_E(u)\;.
\end{equation}
Therefore, the large $\bar k$ WKB analysis proceeds in the same way as for the longitudinal components, and the connection coefficients have the save asymptotic behaviour given in \eqref{eq:WKBalphabeta}.

\begin{figure}
    \centering
    \begin{subfigure}{0.45\textwidth}
        \centering
        \includegraphics[width=\textwidth]{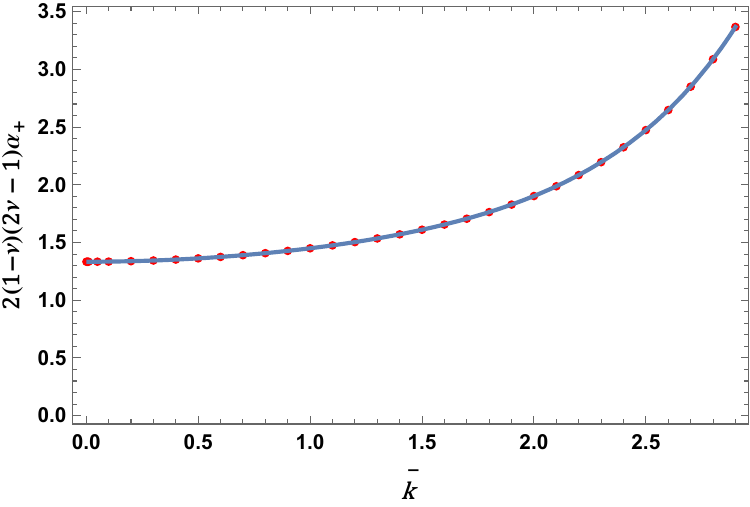}
        \caption{Plot of $2(1-\nu_k)(2\nu_k-1)\alpha_+$ as function of $\bar{k}$.}
        \label{fig:alphaplus}
    \end{subfigure}
    \hfill
    \begin{subfigure}{0.45\textwidth}
        \centering
        \includegraphics[width=\textwidth]{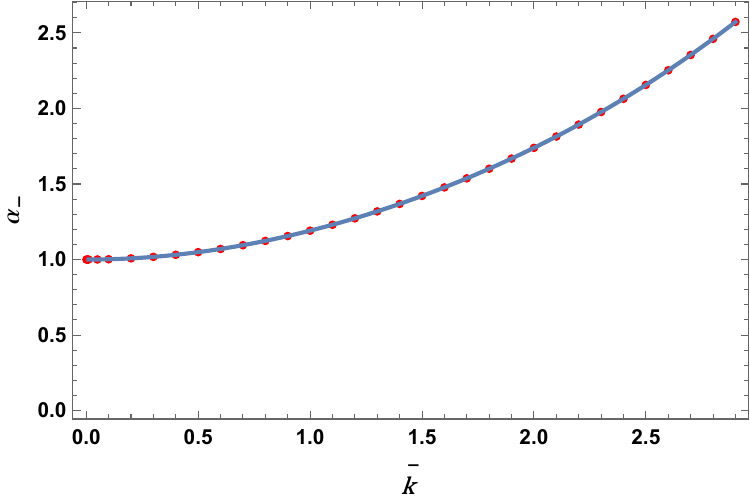}
        \caption{Plot of $\alpha_-$ as function of $\bar{k}$.}
        \label{fig:alphaminus}
    \end{subfigure}
    \caption{Plots of coefficients $\alpha_\pm$ associated to the non-normalizable mode as functions of $\bar{k}$.}
\end{figure}

\begin{figure}
    \centering
    \begin{subfigure}{0.45\textwidth}
        \centering
        \includegraphics[width=\textwidth]{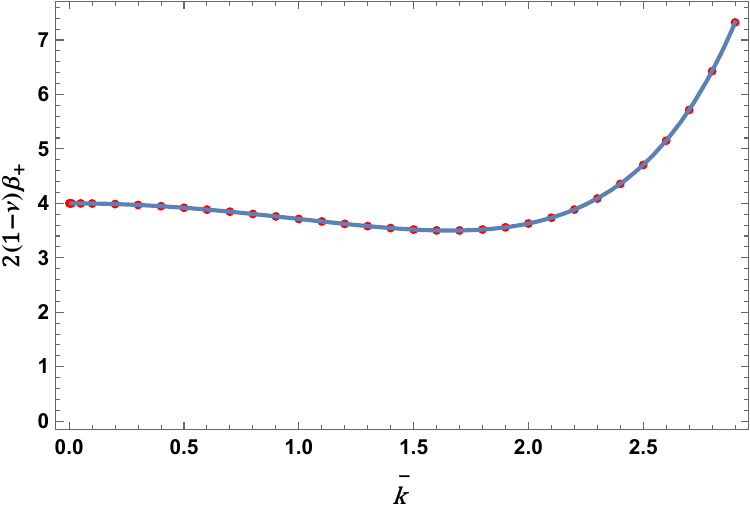}
        \caption{Plot of $2(1-\nu_k)\beta_+$ as function of $\bar{k}$.}
        \label{fig:betaplus}
    \end{subfigure}
    \hfill
    \begin{subfigure}{0.45\textwidth}
        \centering
        \includegraphics[width=\textwidth]{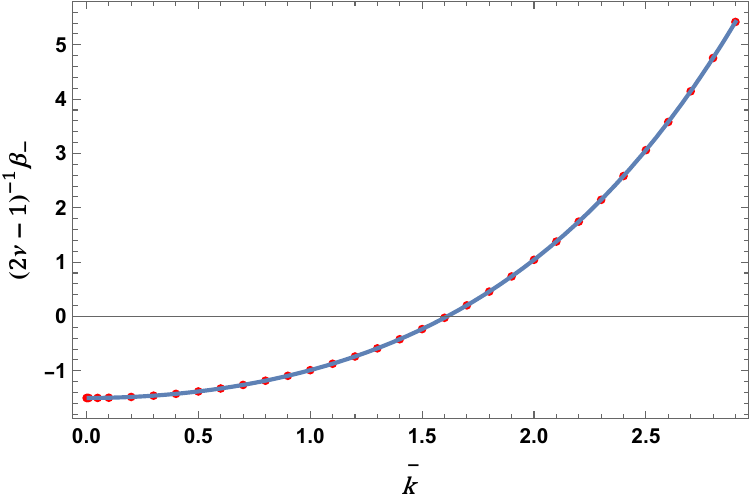}
        \caption{Plot of $(2\nu_k-1)^{-1}\beta_-$ as function of $\bar{k}$.}
        \label{fig:betaminus}
    \end{subfigure}
    \caption{Plots of coefficients $\beta_\pm$ associated to the normalizable mode as functions of $\bar{k}$.}
\end{figure}

We have plotted the connection coefficients in Figures \ref{fig:alphaplus}- \ref{fig:betaminus}. In Figures \ref{fig:alphaplus} and \ref{fig:alphaminus} we plot $\alpha_+$ and $\alpha_-$. The coefficient of the non-normalizable mode, $\alpha_+$, is multiplied by the factor $2(1-\nu_k)(2\nu_k-1)$ in order to remove the divergences in $\bar{k} =0,\,3$. We will examine the origin of this divergence in more detail below. 
In Figures \ref{fig:betaplus} and \ref{fig:betaminus} are shown the plots of $2(1-\nu_k)\beta_+$ and $(2\nu_k-1)^{-1}\beta_-$. 

\subsubsection{Low momentum behavior of connection coefficients}

When $\bar k=0$ ($\nu_k=1/2$) the solutions in the outer region are
\begin{equation}
 \eta_{L,-}^0(Z)=1,\qquad \eta_{L,+}^0(Z)=\frac{2}{1-Z^2}+\frac{4}{3}\log\left[\frac{1+2Z^2}{1-Z^2}\right]-\frac{1}{2}-\frac{4}{3}\log\frac{3}{2}\;.
\end{equation}
The expansions at the boundary and the horizon of the $\eta_{L,+}^0$ solution are
\begin{eqnarray}\label{eq:eta0expMT}
\eta_{L,+}^0(Z) &\sim &\frac{3}{2}-\frac{4}{3}\log\frac{3}{2}+6Z^2+\cdots,\\
\eta_{L,+}^0(Z) &\sim & (1-Z)^{-1}-\frac{4}{3}\log(1-Z)+\cdots\,.
\end{eqnarray}
Therefore, the connection coefficients of these solutions are
\begin{equation}
\alpha_-^0=1,\ \beta_-^0=0,\quad \alpha_+^0=\frac{3}{2}-\frac{4}{3}\log\frac{3}{2},\ \beta_+^0=6.
\end{equation}
We can relate these to the $\bar k\to 0$ limit of the connection coefficients of the non-zero momentum solutions. First, note that the coefficient $j_{-,n}$ in \eqref{eq:coefj} go to zero in the $\bar k\to 0$ limit, so $\eta_{L,-}\to \eta_{L,-}^0$. On the other hand $j_{+,1}$ in \eqref{eq:coefj} is singular in this limit, but the singularity does not propagate to other coefficients in the expansion, one can already see that it is absent in $j_{+,2}$. In order to remove the singular term we can take the following combination of the solutions
\begin{equation}
\tilde \eta_{L,+}(Z)=\eta_{L,+}(Z)-j_{1,+} \eta_{L,-}(Z)\;.
\end{equation}
Expanding this solution at the horizon and for $\bar k\to 0$ one can check that it coincides with the expansion in \eqref{eq:eta0expMT}. We then conclude that  $\tilde \eta_{L,}\to \eta_{L,+}^0$ in the $\bar k\to 0$ limit. This determines the $\bar k\to 0$ behaviour of the connection coefficients
\begin{equation}
\alpha_-\sim \alpha_-^0,\,\,\,\,\, \beta_-\sim \beta_-^0,\qquad \alpha_+\sim \alpha_+^0+j_{+,1}\alpha_-^0,\,\,\,\,\,  \beta_+\sim \beta_+^0+j_{+,1}\beta_-^0\,.
\end{equation}
Since $j_{+,1}\to \infty$ and $\beta_-\to 0$, in order to completely determine $\beta_+$ we need to find the non-zero subleading contribution to $\beta_-$. We expand
\begin{equation}
\eta_{L,-}(Z)= (1-Z)^{-\frac{1}{2}(1-2\nu_k)}\left(1+\left(\nu_k-\frac{1}{2}\right)\eta_-^1(Z)+O\left[\left(\nu_k-\frac{1}{2}\right)^2\right]\right)\;.
\end{equation}
The solution for $\eta_-^1(Z)$ can be found explicitly
\begin{equation}
\begin{split}
\eta_-^1(Z)=&\frac{1}{9} \left(\log (1-Z) (9-12 \log (Z))+9 \log (1+Z)+\pi ^2+9-9 \log (2)-6\text{Li}_2(-2)\right.\\
   &\left.+6 \log (Z) \left(2 \log \left(\frac{1+2
   Z^2}{1+Z}\right)+\frac{3 Z^2}{1-Z^2}\right) + 6\text{Li}_2(-Z^2)-6\text{Li}_2(Z^2)\right)\;.
 \end{split}
\end{equation}
The boundary and horizon expansions are
\begin{eqnarray}\label{eq:eta0exp}
\eta_{L,-}^1(Z) &\sim &\frac{1}{9}\left(9 + \pi^2 - 9\log(2)- 6\text{Li}_2(-2)\right) + 3 Z^2 (-1 +2 \log Z)+\cdots,\\
\eta_{L,-}^1(Z) &\sim & \log(1-Z)+ \frac{11}{6}(1-Z)+\cdots\,.
\end{eqnarray}
Therefore
\begin{equation}
\beta_-\sim -\frac{3}{2}(2\nu_k-1)\;.
\end{equation}
Using now that to leading order
\begin{equation}
j_{+,1}\sim \frac{4}{3(2\nu_k-1)},
\end{equation}
We find that when $\bar k\to 0$
\begin{subequations}
\begin{eqnarray}
&\alpha_-\sim 1+O\left(\nu_k-\dfrac{1}{2}\right),\ &\beta_-\sim -\frac{3}{2}(2\nu_k-1)+O\left[\left(\nu_k-\frac{1}{2}\right)^2\right]\,,\\
&\alpha_+\sim \dfrac{4}{3(2\nu_k-1)}+O(1),\ &\beta_+\sim 4+O\left(\nu_k-\frac{1}{2}\right)\;.
\end{eqnarray}
\end{subequations}

\bibliographystyle{utphys}
\bibliography{biblio}

\end{document}